\documentclass[twocolumn, prl, superscriptaddress,reprint]{revtex4-1}
\usepackage{graphicx}
\usepackage[usenames,dvipsnames]{color}
\usepackage{ulem}
\usepackage{CJK}
\usepackage[colorlinks, linkcolor=blue,anchorcolor=blue,citecolor=blue,urlcolor=blue]{hyperref}
\usepackage{physics}
\usepackage{comment}
\usepackage{titlesec}

\newcommand{\Nit}{$^{14}$N }
\newcommand{\ham}{{\mathcal{H}}}
\newcommand{\sx}{\sigma_x}
\newcommand{\sy}{\sigma_y}
\newcommand{\sz}{\sigma_z}

\usepackage{color,soul}

\begin{document}
\begin{CJK*}{UTF8}{}
\title{Observation of symmetry-protected selection rules in periodically driven quantum systems}

\author{Guoqing Wang \CJKfamily{gbsn}(王国庆)}
\thanks{These authors contributed equally.}
\affiliation{
   Research Laboratory of Electronics and Department of Nuclear Science and Engineering, Massachusetts Institute of Technology, Cambridge, MA 02139, USA}
\author{Changhao Li \CJKfamily{gbsn}(李长昊)}
\thanks{These authors contributed equally.}
\affiliation{
   Research Laboratory of Electronics and Department of Nuclear Science and Engineering, Massachusetts Institute of Technology, Cambridge, MA 02139, USA}

\author{Paola Cappellaro}\email[]{pcappell@mit.edu}
\affiliation{
   Research Laboratory of Electronics and Department of Nuclear Science and Engineering, Massachusetts Institute of Technology, Cambridge, MA 02139, USA}
\affiliation{Department of Physics, Massachusetts Institute of Technology, Cambridge, MA 02139, USA}

\begin{abstract}
Periodically driven quantum systems, known as Floquet systems, have been a focus of non-equilibrium physics in recent years, thanks to their rich dynamics. Not only time-periodic systems exhibit symmetries similar to those in  spatially periodic systems, but they also display novel behavior due to symmetry breaking. Characterizing such dynamical symmetries is crucial, but the task is often challenging, due to limited driving strength and the lack of an experimentally accessible characterization protocol. Here, we show how to  characterize dynamical symmetries including parity, rotation, and particle-hole symmetry by observing the symmetry-induced selection rules between Floquet states.  Specifically, we exploit modulated quantum driving to reach the  strong light-matter coupling regime  and we introduce a protocol to experimentally extract the transition elements between Floquet states from the coherent evolution of the system.  Using the nitrogen-vacancy center in diamond as an experimental testbed, we apply our methods to observe symmetry-protected dark states and dark bands, and the coherent destruction of tunneling effect. Our work shows how to exploit the quantum control toolkit to study dynamical symmetries that can arise in topological phases of strongly-driven Floquet systems. 
\end{abstract}

\maketitle
\end{CJK*}	

\textit{Introduction- }
Intriguing physical phenomena, such as  topological phases, can emerge from symmetries, which play an important role in determining  a system properties~\cite{Kitaev2009,angeli__2021,cao_unconventional_2018,seyler_signatures_2019,tran_evidence_2019,shirleySolutionSchrodingerEquation1965,PRXQuantum.2.020320}.  Symmetries can lead to general universality classes, so that, for example,  phases of topological insulators can be arranged into a periodic table~\cite{Kitaev2009}. 
Engineering novel quantum materials with desired symmetry properties~\cite{angeli__2021,cao_unconventional_2018,seyler_signatures_2019,tran_evidence_2019} can be challenging. Time-periodic systems provide a versatile alternative solution, even allowing to study novel dynamical phases that cannot be realized in static systems~\cite{RoyPRB2017,PengPRL2019,WinterspergerNPhy2020}, as the periodic driving can force the system into topological phases~\cite{RechtsmanNphy2013,GregorNature2014,JimenezPRL2015}.
These  dynamical symmetries in the time domain are described by Floquet theory~\cite{shirleySolutionSchrodingerEquation1965}, in analogy with spatial symmetries described by Bloch theory. 

The dynamical symmetries give rise to transition selection rules between Floquet states that have recently been analyzed theoretically~\cite{engelhardt_dynamical_2021}. 
However, to experimentally observe the transition rules and characterize the symmetries in the time domain is challenging: first, the strong light-matter coupling regime is required to generate high-order Floquet bands, but this is often difficult to reach in practice due to finite  strength of the driving fields. Second, observation of the selection rules requires en experimental toolkit that enables extraction of the transition elements between Floquet states. In this letter, we tackle both challenges and provide a practical solution by combining modulated driving with the observation of the subsequent coherent dynamics of quantum systems to experimentally detect symmetry-induced selection rules -- and their breaking. 

Recent years have witnessed rapid development of exquisite quantum control techniques that can enable engineered driving beyond hardware limitations~\cite{AjoyPNAS2017}. For example, concatenated continuous driving (CCD) is a continuous dynamical decoupling technique introduced to counteract driving inhomogeneities and achieve longer coherence times  ~\cite{caiRobustDynamicalDecoupling2012,farfurnikExperimentalRealizationTimedependent2017,wangCoherenceProtectionDecay2020}. 
More recently, it has been shown to allow reaching the strong-coupling regime even with limited driving fields, while also enabling the observation of phenomana, such as the high-order Mollow triplets beyond the rotating wave approximation~\cite{wang_observation_2021} that are usually ``invisible'' in simple driving protocols. Here we will show that not only we can exploit modulated driving to reach an effective strong-coupling regime, but also to engineer driving transitions (such as double-quantum transitions) that are not directly accessible given the driving field operators.  


To extract the transition elements between Floquet states in the strongly-driven system, we further  develop a protocol based on monitoring the coherent state evolution by projective measurements, which  maps the dynamical dipole matrix elements describing  Floquet band transitions to measurable Rabi oscillation amplitudes.  We take advantage of the controllability and long coherence achieved even in large qubit-like systems, avoiding the need for the  ``pump-probe'' method traditional in atomic and optical physics and for dissipation. Our method,  applicable to general N-level quantum systems, is thus convenient in many modern (many-qubit) quantum systems. 

Exploiting these technical advances, we are able to experimentally study parity and particle-hole symmetry by monitoring the evolution of two levels of a Nitrogen-Vacancy (NV) center in diamond, under modulated driving. Our experiments reveal the emergence of dark states and dark bands --and their disappearance when the symmetries are broken-- as well as  coherent destruction of tunneling,  which corresponds to vanishing transition rates at the degeneracy of quasi-energies. We further show that the modulated driving can engineer a rotationally-symmetric Hamiltonian over the three levels of the NV centers, further indicating that our methods are broadly applicable and an important step toward exploring topological phases that arise in Floquet systems~\cite{RoyPRB2017}.


\textit{Technique to characterize dynamical symmetries- }\\
\textit{Characterization of dynamical symmetries by generalized Rabi oscillations}
It is well-known that spatially periodic Hamiltonians in solid state physics  can be analyzed by Bloch theory,  which predicts a periodic structure in the reciprocal space.
Similarly, the  dynamics of periodically driven qubits with Hamiltonian $\ham(t)=\ham(t+T)$ is solved by  Floquet theory, yielding a series of equidistant energy bands (manifolds) $\lambda^\mu+n\omega_m$ ($n\in Z$) with Floquet eigenenergies $\lambda^\mu$ and frequencies $\omega_m=2\pi/T$~\cite{leskesFloquetTheorySolidstate2010,shirleySolutionSchrodingerEquation1965}. The time-dependent Schr\"{o}dinger equation $i\frac{\partial}{\partial t}\ket{\Psi(t)}=\ham(t)\ket{\Psi(t)}$ is indeed equivalent to the eigenvalue problem for a time-independent Floquet matrix $[\ham(t)-i\frac{\partial}{\partial t}]\ket{\Phi^\mu(t)}=\lambda^\mu\ket{\Phi^\mu(t)}$, with $\ket{\Psi^\mu(t)}=e^{-i\lambda^\mu t}\ket{\Phi^\mu(t)}$. 
The Floquet eigenvectors,  $\ket{\Phi^\mu(t)}=\ket{\Phi^\mu(t+T)}$, have the same period as the Hamiltonian,  so that we can write $\ket{\Phi^\mu(t)}=\sum_{n=-\infty}^{+\infty} e^{-in\omega_m t}\ket{\Phi_n^\mu}$~\cite{leskesFloquetTheorySolidstate2010}. 

Consider a time-independent symmetry operator $\hat{\mathcal{S}}$ (e.g., rotation, parity, particle-hole, etc.) satisfying
\begin{equation}
    \hat{\mathcal{S}}\left[\ham(\beta_\mathcal{S} t+t_\mathcal{S})-i\frac{d}{dt}\right]\hat{\mathcal{S}}^{-1}=\alpha_\mathcal{S}\left[\ham(t)-i\frac{d}{dt}\right],
\end{equation}
where the parameters $\{\alpha_\mathcal{S},\beta_\mathcal{S}\}\in\{1,-1\}$ and $t_\mathcal{S}$ define the details of the symmetry.  
The Floquet eigenstate $\ket{\Phi^\mu(t)}$ also has the same symmetry $\ket{\Phi^{\mu^\prime}(t)}=\pi_\mu^{\mathcal{S}}\hat{\mathcal{S}}\ket{\Phi^\mu(\beta_\mathcal{S} t+t_\mathcal{S})}$, with $|\pi_\mu|=1$,
as derived in Ref.~\cite{engelhardt_dynamical_2021}. 
These symmetries can be probed by evaluating the susceptibility, e.g.  in  light scattering experiments of a probe field $\hat{V}$,  in analogy with  ``pump-probe'' schemes common in atomic and optical physics. 
The susceptibility depends on the dynamical dipole matrix elements associated with the probing operator ${V}$  
\begin{align}
    V_{\mu,\nu}^{(n)}&=\frac{1}{T}\int_0^T \langle\Phi^\mu(t)|{V}|\Phi^\nu(t)\rangle e^{-in\omega_m t}dt, 
\end{align} 
where $n$ denotes the order of the energy band,  where $1/T\int_0^T AB$ defines the scalar product in Floquet space. 
When $\hat{\mathcal{S}}^\dagger{V}\hat{\mathcal{S}}=\alpha_V{V}$ (or $\alpha_V{V}^*$ for particle-hole symmetry), the dynamical symmetry gives rise to symmetry-protected selection rules including symmetry-protected dark states (spDSs) ($V_{\mu,\nu}^{(n)}=0$), symmetry-protected dark bands (spDBs)  (vanishing susceptibility for complete bands)
, and symmetry induced transparency (siT) (destructive interference between non-zero $V_{\mu,\nu}^{(n)}$ elements)~\cite{SOM,engelhardt_dynamical_2021}.

Instead of measuring the susceptibility in a pump-probe experiment as proposed in \cite{engelhardt_dynamical_2021}, here we establish a general experimental method to directly measure the dipole operator $V$. 
Specifically, we draw a correspondence between the transition dipole matrix elements, typical of light scattering experiments, and measurable Rabi oscillation amplitudes arising in the context of coherent state evolution. We show that in a coherent system, the amplitudes of the Fourier components of $\langle V(t)\rangle$ display the desired properties (e.g. spDBs, spDSs, siT, etc.) associated to $V_{\mu\nu}^{(n)}$.

We start by considering a general N-level quantum system. According to  Floquet theory, the coherent state evolution is a superposition of different Floquet eigenstates $\ket{\Psi(t)}=\sum_{\mu}c^{\mu}e^{-i\lambda^\mu t}\ket{\Phi^\mu(t)}$$=\sum_{\mu,n}c^\mu\ket{\Phi_n^\mu}e^{-in\omega_m t}e^{-i\lambda^\mu t}$, where the coefficients $c^\mu$ is given by the initial condition at $t=0$. 
The expectation value of the probing operator in the ``pump-probe'' scheme, $\langle V\rangle=\bra{\Psi(t)}V\ket{\Psi(t)}$ can then be written as 
\begin{equation}
\langle V\rangle = \sum_{\mu,\nu,n}c^{\mu*} c^{\nu} e^{i(\lambda^\mu-\lambda^\nu)t}e^{in\omega_m t}V_{\mu,\nu}^{(n)},
\end{equation}        
where the dynamical dipole matrix element $V_{\mu,\nu}^{(n)}$ can be calculated as
\begin{align}
    V_{\mu,\nu}^{(n)}&=\frac{1}{T}\int_0^T \langle\Phi^\mu(t)|{V}|\Phi^\nu(t)\rangle e^{-in\omega t}dt\nonumber\\
    &=\sum_k \mathcal V_k\sum_p\bra{\Phi_p^\mu}k\rangle\langle k\ket{\Phi_{p-n}^\nu}.
\end{align}
Here we introduced the spectral decomposition of $V$, ${V}=\sum_{k} \mathcal{V}_k\ket{k}\bra{k}$. By considering the Fourier decomposition of $\langle V\rangle$,  
\begin{equation}
\label{Eq_P0_appendix}
  \langle V\rangle  = \sum_{\mu,\nu,n} |A_{\mu,\nu}^{(n)}|\cos(\omega_{\mu,\nu}^{(n)} t+\phi_{\mu,\nu}^{(n)}),
\end{equation}        
with frequencies $\omega_{\mu,\nu}^{(n)}=n\omega_m+(\lambda^\mu-\lambda^\nu)$, we find that the Fourier amplitudes 
\begin{equation}
    A_{\mu,\nu}^{(n)} = |A_{\mu,\nu}^{(n)}|\exp(i\phi_{\mu,\nu}^{(n)})=2c^{\mu*}c^{\nu}
V_{\mu,\nu}^{(n)}
\end{equation}
can be used to extract the dynamical dipole matrix elements.

We note that sometimes the operator $V$ cannot be directly measured. In that case, we can use control on the system preparation and readout to separately monitor the overlap of the state with eigenstates of $V$, $P_{\ket{k}}(t)=|\bra k\Psi(t)\rangle|^2$. We can then analyze the  ``weighted Rabi'' amplitude 
\begin{equation}
    P(t)=\sum_k \frac{\mathcal{V}_k}{\mathcal{V}} P_{\ket{k}}(t)\equiv \frac{\langle{V}\rangle}{\mathcal{V}}
\end{equation}
where $\mathcal{V}=\sum_k |\mathcal{V}_k|$ is a normalization factor. 
The weighted Rabi oscillations can then be decomposed to a series of frequency components with amplitudes $a_{\mu,\nu}^{(n)}=A_{\mu,\nu}^{(n)}/\mathcal V$, which can be used to investigate symmetry properties.
Consider for example a two-level system (TLS). The probing operator $V$ is then a combination of Pauli operators $\sigma_j$ with eigenvectors $\ket{0_j}, \ket{1_j}$ and normalized eigenvalues $\pm1$.  The weighted Rabi oscillations have the form $P(t)=(1/2)\left[P_{\ket{0_j}}(t)-P_{\ket{1_j}}(t)\right]$ that can be simplified to the typical Rabi oscillations $P(t)+1/2=P_{\ket{0_j}}(t)$, thus clarifying the connection of our protocol with typical Rabi measurements. 

In addition to using control of the readout state to measure $P_{\ket k}$, we can also control the initial state to extract information about selected dipole matrix elements, by appropriately choosing the coefficients $c^\mu$. 

When $\mu=\nu$, all bands under the same order $(n)$ share the same frequency $n\omega_m$  (\textit{centerbands}), and the observed Rabi component is their coherent interference with an amplitude $a_{0}^{(n)} = |a_{0}^{(n)}|\exp(i\phi_{0}^{(n)})=2\sum_{\mu} |c^{\mu}|^2 V_{\mu,\mu}^{(n)}/\mathcal{V}$. 
Although the measured centerbands are indeed the interference between many degenerate bands, we can observe each band $V_{\mu,\mu}^{(n)}$ individually by tuning the initial condition $|c^\mu|=1$ (\textit{quantum mode control})~\cite{wang_observation_2021}. 

When $\mu\neq\nu$, the off-diagonal dynamical dipole matrix elements $V_{\mu,\nu}^{(n)}$ can be mapped to the Rabi amplitudes $a_{\mu,\nu}^{(n)}$ corresponding to the bands $n\omega_m+(\lambda^\mu-\lambda^\nu)$ (\textit{sidebands}). At the degeneracy points when $\lambda^\mu=\lambda^\nu$, different sidebands $a_{\mu,\nu}^{(n)},a_{\nu,\mu}^{(n)}$ interfere with each other,  inducing phenomena such as the siT, or more generally the Landau-Zener-St\"{u}ckelberg interferometry~\cite{shevchenkoLandauZenerStuckelberg2010,huangLandauZenerStuckelbergInterferometrySingle2011} and coherence destruction of tunneling (CDT)~\cite{grossmannCoherentDestructionTunneling1991,grossmannCoherentTransportPeriodically1993,LuBlochSiegertPhysRevA.86.023831,zhouObservationTimeDomainRabi2014}.


\begin{table}[htbp]
\caption{Correspondence between the measurable Rabi amplitudes $a_{i}^{(n)}$ and dipole matrix element $V_{\mu,\nu}^{(n)}$ {for a TLS}. 
Note that $\Phi_{p,0_j}^{\mu}=\langle 0_j\ket{\Phi_{p}^\mu}$, and listed cases do not include $i=n=0$~\cite{wang_observation_2021,SOM}.}
\label{Correspondence}
\begin{tabular}{c|c|c}
\hline\hline
Bands$_{\{i,n\}}$ &  Rabi amplitudes $a_{\mu,\nu}^{(n)}$ & Expressed in $V_{\mu,\nu}^{(n)}$\\\hline
\hline
\{0,n\} & $2\sum_{\pm}|c^\pm|^2\sum_k \Phi_{k+n,0_j}^{\pm*}\Phi_{k,0_j}^{\pm}$ & $|c^+|^2V_{+,+}^{(n)}+|c^-|^2V_{-,-}^{(n)}$ \\
\hline
\{-1,n\}  & $2\sum_k c^{+}c^{-*}\Phi_{k,0_j}^{+}\Phi_{k+n,0_j}^{-*}$ &$c^+c^{-*}V_{-,+}^{(n)}$ \\
\hline
\{+1,n\}  & $2\sum_k c^{+*}c^{-}\Phi_{k+n,0_j}^{+*}\Phi_{k,0_j}^{-}$ &$c^{+*}c^{-}V_{+,-}^{(n)}$  \\
\hline\hline
\end{tabular}
\end{table}

\textit{Applications - }
To demonstrate the power of combining modulated driving with weighted Rabi measurements we characterize symmetries arising in two- and three-level systems, that we can experimentally realize using NV centers in diamond.
NV centers are atom-like solid-state defects with a triplet ground state labelled by $\ket{m_s=0,\pm1}$ with long-coherence time that enables their application in broad field of quantum information science, including quantum sensing~\cite{DegenRMP2017,SchirhaglReview2014,LiNL2019,SchmittScience2017,wang2021nanoscale} and quantum control~\cite{RechtsmanNphy2013,wangCoherenceProtectionDecay2020}.

For a two-level system (TLS) we use $\mu\in\{+,-\}$ to denote the two Floquet eigenstates, and decompose the Rabi oscillation $P_{\ket{0_j}}(t)$ (where $j=\{x,y,z\}$) into a series of Mollow triplets $\omega_{i}^{(n)}=n\omega_m+i(\lambda^+-\lambda^-)$.  Here $i=0,\pm 1$ correspond to the centerbands and sidebands, respectively, with amplitudes $a_{\pm 1}^{(n)} = |a_{\pm 1}^{(n)}|\exp(i\phi_{\pm 1}^{(n)})=c^{\pm*}c^{\mp}V_{\pm,\mp}^{(n)}$, $a_{0}^{(n)}=\sum_{\pm}|c^\pm|^2 V_{\pm,\pm}^{(n)}$~\cite{wang_observation_2021}. The correspondence between the dipole matrix element $V_{\mu,\nu}^{(n)}$ and the Rabi amplitudes $a_{i}^{(n)}$ is listed in Table~\ref{Correspondence}.

To reduce the 3-level NV center to an effective TLS,  we break the $\ket{m_s=\pm1}$ degeneracy by applying  an external magnetic field with strength 239 Gauss in our setup~\cite{jaskulaPhysRevApplied.11.054010,wang_observation_2021,wangCoherenceProtectionDecay2020}. Then, the $\ket{m_s=0}$ and $\ket{m_s=-1}$ states form a TLS which is driven by resonant microwave.
We simultaneously address an ensemble of non-interacting NV centers ($\sim10^{10}$ qubits) to increase the signal-to-noise ratio and an arbitrary waveform generator (WX1284C) is used to generate the desired waveform for Hamiltonian engineering.

To  engineer strong driving on the NV centers, we rely on the phase-modulated  CCD driving technique~\footnote{Here we did not choose  amplitude-modulated CCD, since it is still limited by the total microwave power and amplifier nonlinearity~\cite{wang_observation_2021}.}, which has  been applied before to approach the strong-coupling regime even with limited driving fields~\cite{caiRobustDynamicalDecoupling2012,farfurnikExperimentalRealizationTimedependent2017,wangCoherenceProtectionDecay2020}. 
We apply a phase-modulated waveform   
\begin{equation}
    \ham=\frac{\omega_0}{2}\sz+\Omega\cos\left(\omega t+\frac{2\epsilon_m}{\omega_m}\cos(\omega_m t+\phi)\right)\sx,
    \label{Hamiltonian_PhaseMod}
\end{equation}
where $\omega=\omega_0=(2\pi)2.20\text{GHz}$ is the qubit frequency, $\Omega$  the microwave driving strength, and $\epsilon_m,\omega_m,\phi$ are modulation parameters.
In the interaction picture defined by $U=\exp\left[-i\left(\frac{\omega t}{2}\sz+\epsilon_m \frac{\cos(\omega_m t+\phi)}{\omega_m}\sz \right)\right]$, the Hamiltonian  $\ham_I=U^\dagger \ham U-U^\dagger i\frac{d}{dt}U$  is
\begin{equation}
        \ham_I=\frac{\Omega}{2}\sx+\epsilon_m\sin(\omega_mt+\phi)\sz.
    \label{HI_PhaseMod}
\end{equation}
We thus obtain a time-periodic Hamiltonian $\ham_I(t)=\ham_I(t+T)$  with $T=(2\pi)/\omega_m$, where $\Omega$($\epsilon_m$) plays the role of a static (driving) field, and their relative strength can be  easily tuned to approach the strong coupling regime, without hardware limitations.

\begin{figure*}[htbp]
\includegraphics[width=0.75\textwidth]{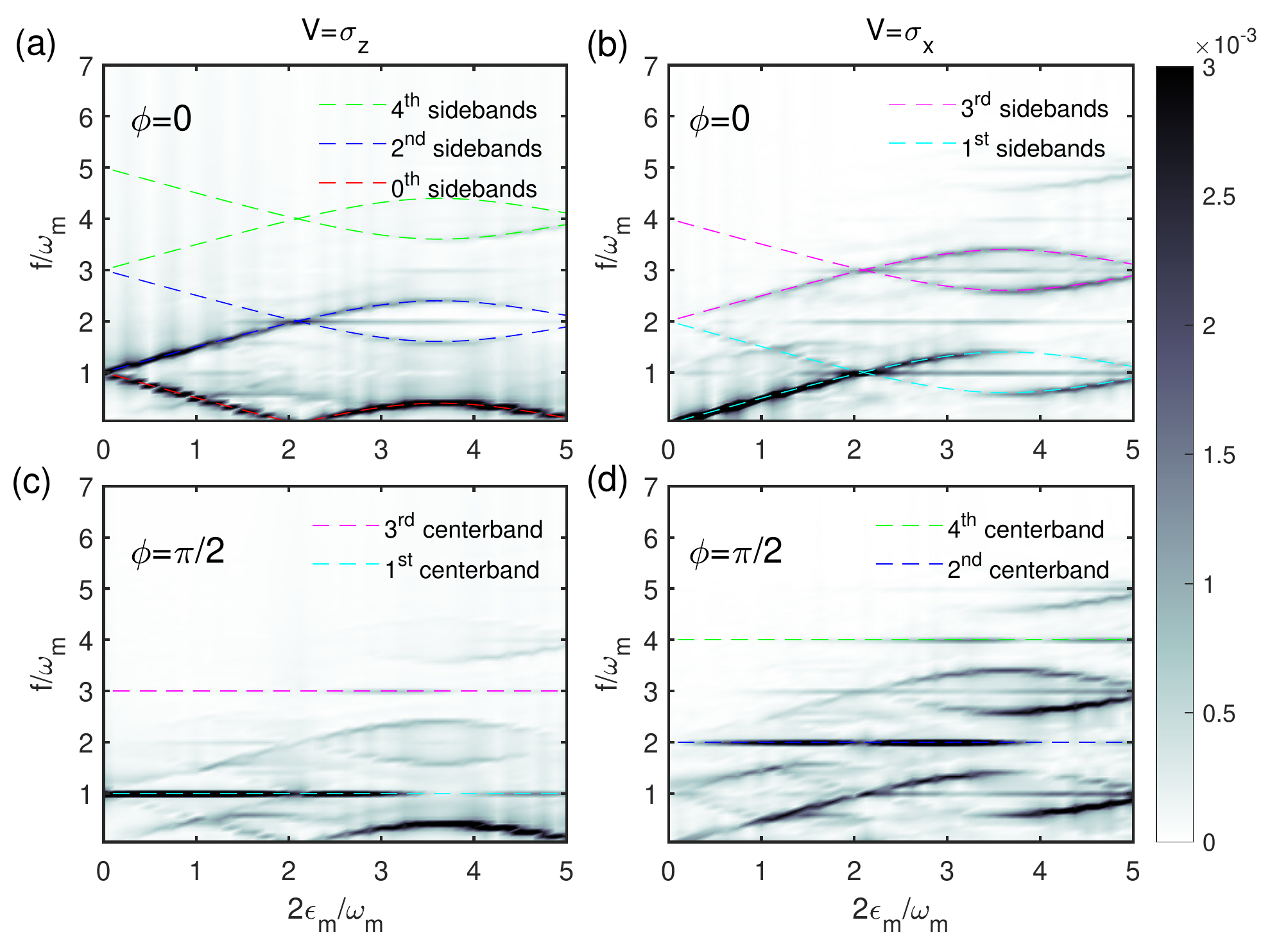}
\caption{\label{Resonance_data} Observation of spDSs, spDBs, and destructive interference between centerbands. (a,c) Fourier spectrum of Rabi measurement $P_{\ket{0}}(t)$ (${V}=\sz$) under different modulation strength $2\epsilon_m/\omega_m$. Parameters are $\Omega=\omega_m=(2\pi)3\text{MHz}$, $\phi=0,\pi/2$ corresponding to (a), (c) respectively. The population is measured from $t=0$ to $t=4\mu$s with 401 sampling points. (b,d) Fourier spectrum of Rabi measurement $P_{\ket{+}}(t)$ (${V}=\sx$) under different modulation strength $2\epsilon_m/\omega_m$. Other parameters are the same as in (a,c).}
\end{figure*}

The periodic Hamiltonian $\ham_I(t)$ in Eq.~\eqref{HI_PhaseMod} has two nontrivial eigenenergies $\lambda^\pm$, with corresponding time-periodic eigenstates $\ket{\Phi^\pm(t)}$, in  the first ``Brillouin zone'' $(-\omega_m/2,\omega_m/2]$ of the Floquet matrix (the complete energy structure is $E=n\omega_m+\lambda^\pm$, with $n$ integer). 
The energy transitions  form the well-known Mollow triplets,  with $n\omega_m$($n\omega_m\pm(\lambda^+-\lambda^-)$) the frequencies of $n^{th}$ centerband (sidebands).
These transitions  can be probed either through  conventional ``pump-probe" spectroscopy, such as spontaneous emission~\cite{mollowPowerSpectrumLight1969}, or via projective Rabi measurements in the context of coherent state evolution~\cite{wang_observation_2021,rohrSynchronizingDynamicsSingle2014,pigeauObservationPhononicMollow2015}.

In the following we experimentally evaluate the dynamical symmetries of the qubit Hamiltonian $\ham_I$, and study the associated spDSs, spDBs, and siT through intensities of the Floquet state transitions (i.e., Mollow triplets),  extracted from Rabi oscillations. 

The first dynamical symmetry  is a 2-fold rotation or parity symmetry $\hat{R}=\sx$, which satisfies $\hat{R}\ham_I(t+T/2)\hat{R}^\dagger=\ham_I(t)$. This  symmetry yields the eigenstate relation $\ket{\Phi^\mu(t)}=e^{i\pi m_\mu}\hat{R}\ket{\Phi^\mu(t+T/2)}$ with $\mu\in\{+,-\}$ and $m_\mu\in\{0,1\}$. The dynamical dipole matrix elements are
$V_{\mu,\nu}^{(n)}\propto\left[1+e^{i\pi(m_\mu-m_\nu)-i\pi n}\alpha_V\right]$, 
where $\alpha_V$ is given by $\hat{R}{V}\hat{R}^\dagger=\alpha_V{V}$.
Since the observation operators  ${V}=\sy,\sz$  have $\alpha_V=-1$,  the  Mollow centerbands (sidebands) with odd (even) orders are visible and the even (odd) orders vanish. The opposite holds for the case of ${V}=\sx$, which sets  $\alpha_V=1$. As a result, a series of spDSs and spDBs are predicted by the parity symmetry.

To experimentally observe the selection rules, we measure the Rabi oscillations of an initial state $\ket{0}$ under different modulation strength $2\epsilon_m/\omega_m$. 
We choose the resonance condition $\Omega=\omega_m$, such that the Mollow bands are linearly dependent on the driving strength $\epsilon_m$ when the rotating wave approximation is valid ($\epsilon_m\ll\Omega$), as shown in Fig.~\ref{Resonance_data}. 
In Figs.~\ref{Resonance_data}(a,c), for the observation operator ${V}=\sz$, the $1^{st},3^{rd}$ Mollow centerbands and $0^{th},2^{nd},4^{th}$ Mollow sidebands are visible while the other orders are vanishing. In Figs.~\ref{Resonance_data}(b,d), with the observation operator ${V}=\sx$, we observe the opposite behavior, as expected.
Note that some unexpected bands (e.g., odd order centerbands in the range of $2.5<2\epsilon_m/\omega_m<5$ in Figs.~\ref{Resonance_data}(b,d)) are still visible, albeit with small intensities. We attribute their appearance to imperfect experimental control such as driving inhomogeneities, that introduce detuning term $(\delta/2)\sz$ in the Hamiltonian $\ham_I$ in Eq.~\eqref{HI_PhaseMod}~\cite{SOM,wangCoherenceProtectionDecay2020}.  

The second symmetry is the particle-hole symmetry $\hat{P_1}=\sz$, which satisfies $\hat{P_1}\ham_I(t+T/2)\hat{P_1}^\dagger=-\ham_I(t)$. Given an eigenstate $\ket{\Phi^\mu(t)}$ corresponding to the eigenenergy $\lambda^\mu$, $\hat{P_1}\ket{\Phi^{\mu *}(t+T/2)}$ is also an eigenstate of the system with eigenenergy $-\lambda^\mu$. 
For our Hamiltonian $\ham_I$ in Eq.~\eqref{HI_PhaseMod}, this yields $\ket{\Phi^{-\mu}(t)}=\pi_\mu^{(P_1)}\hat{P_1}\ket{\Phi^{\mu *}(t+T/2)}$.
The dynamical dipole matrix element is $V_{\mu,\nu}^{(n)}=\alpha_V^{(P_1)}e^{i\pi n}V_{\nu^\prime,\mu^\prime}^{(n)}$, 
where $\alpha_V^{(P_1)}$ is given by the relation $\hat{P_1}^\dagger{V}\hat{P_1}=\alpha_V^{(P_1)}{V}^*$. The selection rules and coherent interference induced by the particle-hole symmetry fall into two cases. 

(1) When $\mu\neq\nu$ (implying $V_{\mu,\nu}^{(n)}=\alpha_V^{(P_1)}e^{i\pi n}V_{\mu,\nu}^{(n)}$), sidebands vanish when $\alpha_V^{(P_1)}e^{i\pi n}=0$. For ${V}=\sz$, $\alpha_V^{(P_1)}=1$, Mollow sidebands with odd order $(n)$ vanish (spDSs, spDBs).  For ${V}=\sx$, $\alpha_V^{(P_1)}=-1$, Mollow sidebands with even order vanish (spDSs, spDBs). These predictions  are consistent with  parity symmetry, and are observed experimentally in Fig.~\ref{Resonance_data}. 

(2) When $\mu=\nu$ (and $V_{+,+}^{(n)}=\alpha_V^{(P_1)}e^{i\pi n}V_{-,-}^{(n)}$), destructive (or constructive) interference is induced when $\alpha_V^{(P_1)}e^{i\pi n}=-1$ (or $\alpha_V^{(P_1)}e^{i\pi n}=1$) under the condition $|c^\pm|^2=1/2$. This property is used in the \textit{quantum mode control}  proposed in Ref.~\cite{wang_observation_2021}. 
When the qubit initial state is one of the Floquet eigenstates, the sidebands vanish since $c^+c^-=0$. When the qubit initial state is an equal superposition of two Floquet eigenstates, a destructive interference may happen in the centerbands and only sidebands are present. Under the modulation phase $\phi=0$ in Figs.~\ref{Resonance_data}(a,b), the Floquet eigenstates are always in $x-y$ plane of the Bloch sphere such that $|c^\pm|^2=1/2$ for the initial state $\ket{\Psi(0)}=\ket{0}$, and the destructive interference happens in the odd (even) centerband when ${V}=\sz$ (${V}=\sx$). Combining with the parity symmetry that makes the opposite orders of centerband vanish, all centerbands vanish under $\phi=0$ as observed in Figs.~\ref{Resonance_data}(a,b). However, under the modulation phase $\phi=\pi/2$, the Floquet eigenstates are in the $x-z$ plane of the Bloch sphere and the condition $|c^\pm|^2=1/2$ is no longer satisfied. The symmetry-allowed centerband appears as shown in Figs.~\ref{Resonance_data}(c,d) and vanishes at $2\epsilon_m/\omega_m\approx3.8$ where $|c^\pm|^2=1/2$ is accidentally satisfied.

To further demonstrate the symmetry-induced selection rules, we break both the parity  and particle-hole symmetries by introducing an additional term in the engineered Hamiltonian,  $\ham'_I=(\Omega/2)\sx+\epsilon_m\sin(\omega_m t)\sz+0.2\epsilon_m\sin(2\omega_mt)\sz$, and measure the same Rabi spectrum again in Fig.~\ref{ResonanceSB_data}. 
The gradual appearance of the symmetry-protected dark bands as $\epsilon_m$   increases clearly signals the symmetry breakdown.

Another type of destructive interference, siT, is observable when two discrete particle-hole symmetries exist in the system and two sidebands become degenerate at $\lambda^\pm=0$.
In the strong coupling and far off-resonance regime  ($\Omega\ll\epsilon_m,\omega_m$), an additional particle-hole symmetry $\hat{P_2}={I}$ arises such that $I\ham_I(t+T/2)I=-\ham_I(t)$. 
The dynamical dipole matrix elements satisfy $V_{\mu,\nu}^{(n)}=\alpha_V^{(P_1^*)} e^{i\pi n}V_{\mu^\prime,\nu^\prime}^{(n)}$, where $\alpha_V^{(P_1^*)}$ is such that $\alpha_V^{(P_1^*)}{V}=\hat{P_1}^{\dagger*}{V}\hat{P_1}^{*}$. When $\alpha_V^{(P_1^*)} e^{i\pi n}=-1$, two sidebands interfere in a destructive way and a siT happens under the initial condition $c^+c^{-*}=c^{+*}c^{-}$. Under the siT condition, the qubit state evolution is suppressed in the direction of the driving field (i.e., the CDT effect, which has been observed both numerically~\cite{LuBlochSiegertPhysRevA.86.023831} and experimentally~\cite{zhouObservationTimeDomainRabi2014}.) We engineer a strong-coupling Hamiltonian with $\Omega=\omega_m/10$, and measure the Rabi spectrum  (Fig.~\ref{CDT_data}.) A siT is observed when two eigenenergies degenerate at $2\epsilon_m/\omega_m=2.4048$, and accidental dark states are observed at $2\epsilon_m/\omega_m\approx 3.8$. In addition, spDSs (or spDBs) are observed as shown by the dashed lines in Fig.~\ref{CDT_data}, similar to Fig.~\ref{Resonance_data}. 
Details on the observation of a constructive interference ($c^+c^{-*}=-c^{+*}c^{-}$)  can be found in~\cite{SOM}.

\begin{figure}[tb]
\includegraphics[width=0.5\textwidth]{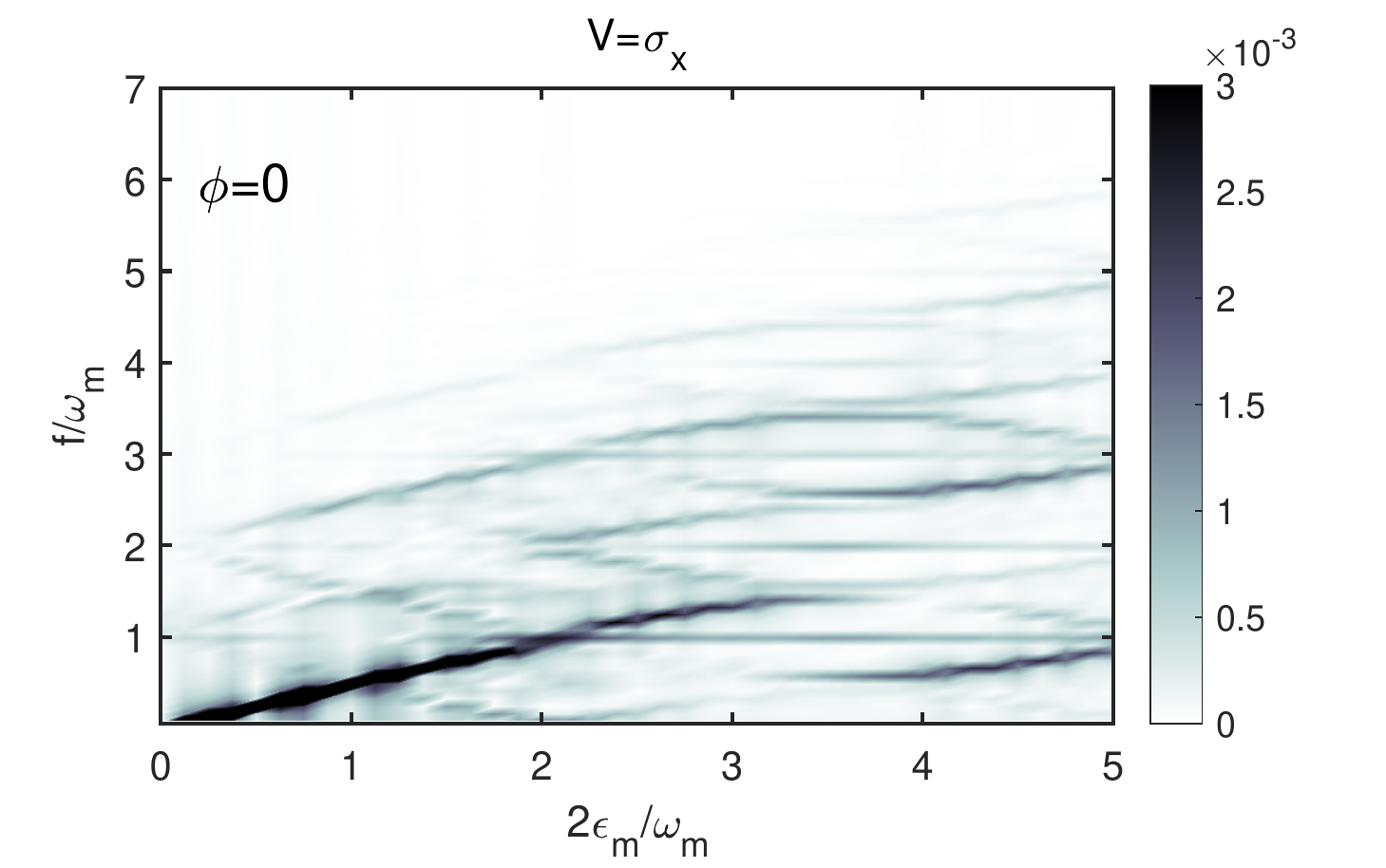}
\caption{\label{ResonanceSB_data} Observation of symmetry breaking. The experimental parameters are the same as in Fig.~\ref{Resonance_data}(b) except for the engineered Hamiltonian $\ham_I=(\Omega/2)\sx+\epsilon_m\sin(\omega_m t)\sz+0.2\epsilon_m\sin(2\omega_mt)\sz$ where the third term oscillating with frequency $2\omega_m$ breaks both two symmetries.}
\end{figure}

\begin{figure}[tb]
\includegraphics[width=0.5\textwidth]{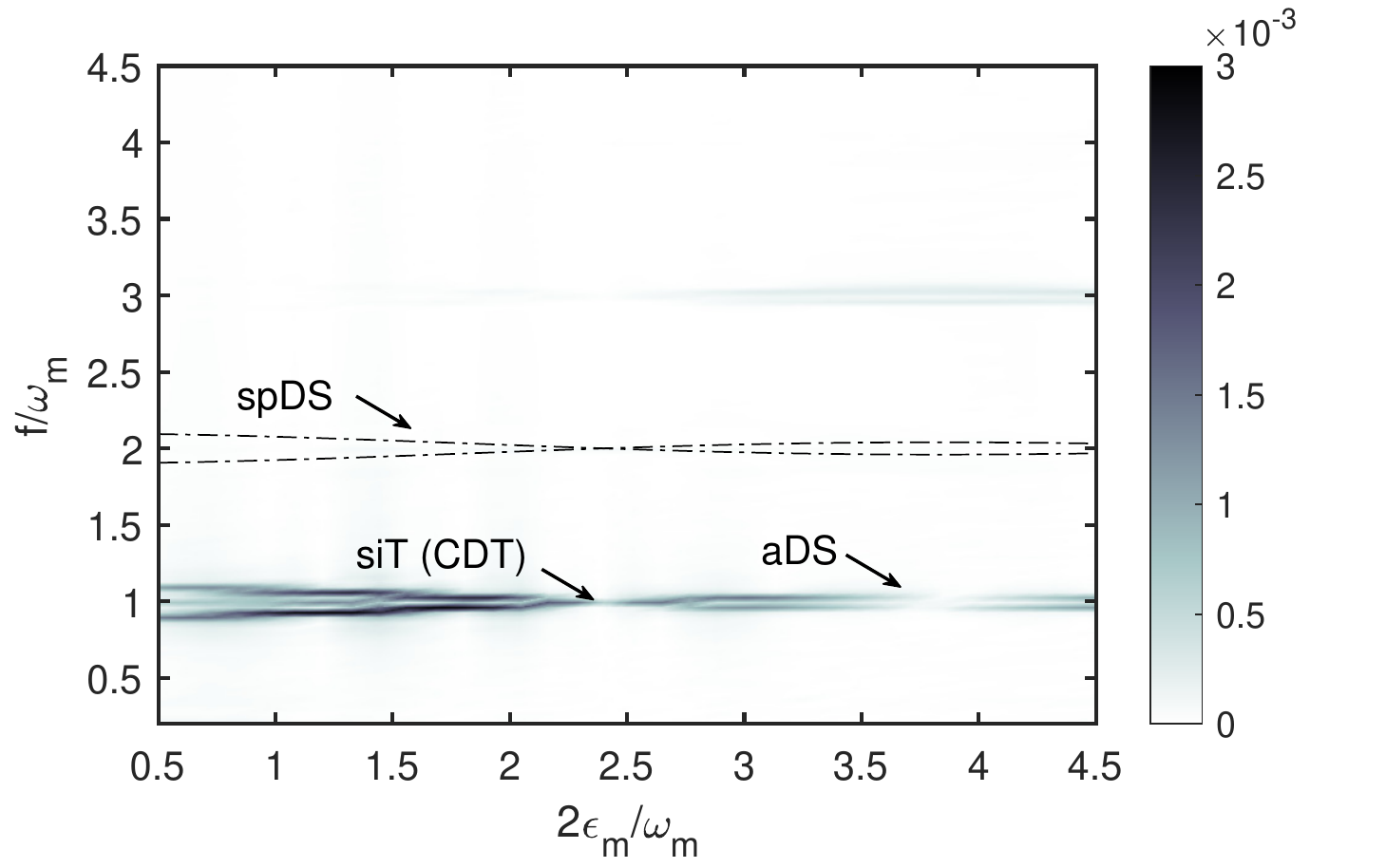}
\caption{\label{CDT_data} Observations of siT (destructive interference between sidebands), spDSs, spDBs and accidental dark states (aDS). Qubit initial state is prepared to $\ket{0}$ and the time-dependent population $P_{\ket{+}}(t)$ is readout (${V}=\sx$) from $t=0$ to $t=2\mu $s with 401 sampling points. Driving parameters are $\Omega=(2\pi)1.5\text{MHz}$, $\omega_m=(2\pi)15\text{MHz}$, $\phi=0$. The intensity plot is the Fourier spectrum of measured Rabi oscillations under different $\epsilon_m$.}
\end{figure}

In order to demonstrate that our technique can be extended beyond TLSs, we show how to use the 3 levels associated with the spin-1 of NV centers  to explore a 3-fold rotation symmetry. We use modulated driving to both reach the strong driving regime and to engineer the double quantum (DQ) transition ($\ket{m_S=-1}\leftrightarrow\ket{m_S=+1}$) in the rotating frame. Indeed, the DQ transition  cannot be directly generated by microwave driving (although it could be achieved by mechanical oscillations~\cite{macquarrieCoherentControlNitrogenvacancy2015,macquarrieMechanicalSpinControl2013a}.) Here we overcome this limitation by simultaneous applying two modulated driving on the single quantum transitions ($\ket{m_S=0}\leftrightarrow\ket{m_S=\pm 1}$), leading to the rotating-frame Hamiltonian~\cite{SOM}
\begin{align}
&\ham^3_I(t)=J\left[\cos(\omega_m t)\ket{-1}\!\bra{+1}+\cos(\omega_m t+2\pi/3)\ket{+1}\!\bra{0}\right. \nonumber
\\
&\left.\qquad\quad +\cos(\omega_m t+4\pi/3)\ket{-1}\!\bra{0}+h.c.\right]	\end{align}
$\ham^3_I(t)$ satisfies a 3-fold rotation symmetry such that $\hat{R}\ham^3_I(t+T/3)\hat{R}^\dagger=\ham_I(t)$ where the rotation operator is $\hat{R}=\ket{-1}\bra{0}+\ket{0}\bra{+1}+\ket{+1}\bra{-1}$. 
Similar to the symmetries discussed above, we find symmetry-protected selections rules by evaluating the Floquet eigenstates and observation operator~\cite{SOM}.
In Fig.~\ref{Spin1_simulation}, we simulate the Fourier spectrum of the weighted Rabi signal $P(t)=(1/4)[2P_{\ket{e_1}}(t)-P_{\ket{e_2}}(t)-P_{\ket{e_3}}(t)]$, which clearly presents the spDSs and spDBs protected by the 3-fold rotation symmetry, as expected. 
\begin{figure}[htbp]
\includegraphics[width=0.5\textwidth]{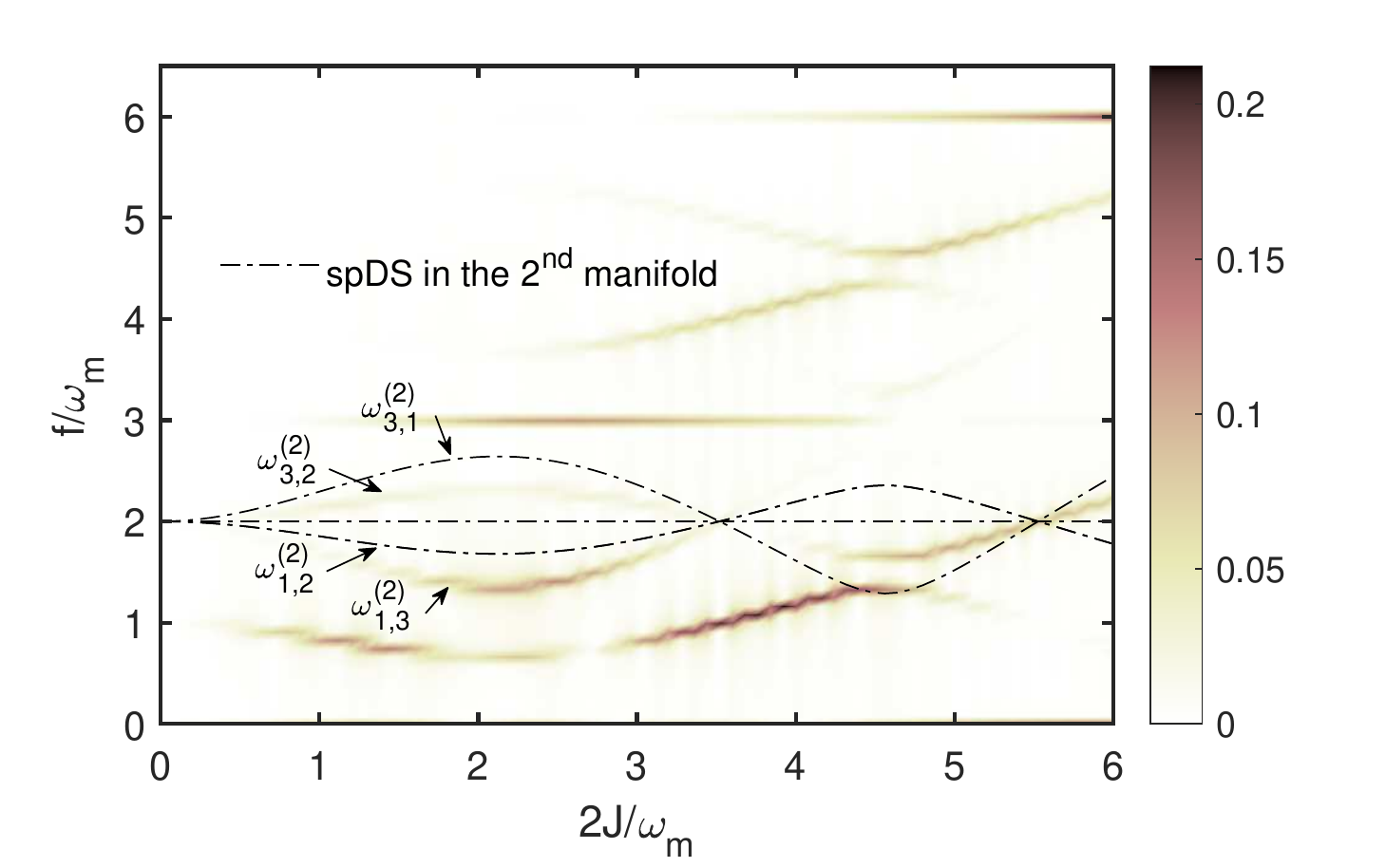}
\caption{\label{Spin1_simulation} Simulation of spDSs and spDBs in a 3-level system. The initial state is prepared to $(1/\sqrt{3})(\ket{e_1}+\ket{e_2}+\ket{e_3})$ where $\ket{e_1}=(1/\sqrt{3})[1,1,1]^T$, $\ket{e_2}=(1/\sqrt{6})[-2,1,1]^T$, $\ket{e_3}=(1/\sqrt{2})[0,1,-1]^T$ are three eigenstates of ${V}$, such that evolution mode involves all bands. The time-dependent weighted Rabi population $P(t)$ is simulated from $t=0$ to $t=40\mu $s with 5001 sampling points. Driving parameters $\omega_m=(2\pi)0.3\text{MHz}$, and $2J/\omega_m$ is swept from $0$ to $6$. The intensity plot is the Fourier spectrum of simulated Rabi oscillations under different $2J/\omega_m$. To avoid confusion, we only show the spDSs (spDBs) in the $2^{nd}$ frequency manifold protected by the rotation symmetry. Please see details in supplement~\cite{SOM}.}
\end{figure}

\textit{Discussions and conclusion- }
By combining modulated driving and detection via Rabi oscillations, we were able to experimentally observe  selection rules protected by dynamical symmetries in a periodic driven solid-state system. 
The modulated driving scheme is instrumental to reach the strong light-matter coupling regime used to reveal high-order Floquet bands; it also introduces additional flexibility in quantum control, enabling to engineer transitions forbidden in the un-modulated frame and revealing details of the dynamics (e.g., Mollow triplets) via mode control. Direct measurement of the dipolar transition operator, $V$ or indirectly via weighted Rabi oscillations is a more efficient strategy than previous ``pump-probe'' methods in the highly coherent quantum systems that can be now routinely engineer.     
Thanks to these techniques, we characterized the parity and particle-hole symmetry in the time domain as well as the CDT effect in the engineered system. In the favor of quantum control, the dynamical symmetries studied here have applications in inducing selections rules for high harmonic generation and driving quantum synchronization~\cite{TindallNJP2020,alon_dynamical_2002,alon_high_2000,alon_selection_1998}.

While we showed simulations and experiments for  two- and three-level systems, the experimental techniques we introduced can be generalized to many-body (N-level) systems in a broad set of  platforms beyond spins, such as cold atoms and superconducting circuits.

When combined with spatial symmetries, dynamical symmetries characterized in this work can lead to novel Floquet topological phases such as Floquet topological insulators and superconductors~\cite{RoyPRB2017}.
The breaking of these dynamical symmetries might lead to intriguing dynamical phase transitions. 
Furthermore, by engineering the dissipation, for example, tuning the decoherence rate of the system, the work here paves the way to explore non-Hermitian Floquet Hamiltonians.

This work was in part supported by ARO grant W911NF-11-1-0400 and NSF grants PHY1734011 and EECS1702716. We thank
Pai Peng for fruitful discussions.

\bibliographystyle{apsrev4-1}
\bibliography{main_text} 
\clearpage

\pagebreak
\widetext
\setcounter{section}{0}
\setcounter{equation}{0}
\setcounter{figure}{0}
\setcounter{table}{0}
\setcounter{page}{1}
\makeatletter
\renewcommand{\theequation}{S\arabic{equation}}
\renewcommand{\thefigure}{S\arabic{figure}}
\renewcommand{\thesection}{S\arabic{section}}


\begin{CJK*}{UTF8}{}
\title{Supplemental materials: Observation of symmetry-protected selection rules in periodically driven quantum systems}

\author{Guoqing Wang \CJKfamily{gbsn}(王国庆)}
\thanks{These authors contributed equally.}
\affiliation{
   Research Laboratory of Electronics and Department of Nuclear Science and Engineering, Massachusetts Institute of Technology, Cambridge, MA 02139, USA}
\author{Changhao Li \CJKfamily{gbsn}(李长昊)}
\thanks{These authors contributed equally.}
\affiliation{
   Research Laboratory of Electronics and Department of Nuclear Science and Engineering, Massachusetts Institute of Technology, Cambridge, MA 02139, USA}

\author{Paola Cappellaro}\email[]{pcappell@mit.edu}
\affiliation{
   Research Laboratory of Electronics and Department of Nuclear Science and Engineering, Massachusetts Institute of Technology, Cambridge, MA 02139, USA}
\affiliation{Department of Physics, Massachusetts Institute of Technology, Cambridge, MA 02139, USA}

\maketitle
\end{CJK*}	
\newpage
\begin{widetext}

\tableofcontents
\clearpage
\section{S1. Dynamics of a periodically driven quantum system}
In this section we introduce the basics of Floquet theory as well as the concatenated continuous driving technique. While much of the introductions and results below have been presented in reference~\cite{wang_observation_2021}, we briefly repeat them here for the convenience of readers.

\subsection{A. Floquet theory}
\label{Derivation: Floquet}
\subsubsection{1. General derivation}
We first give the detailed steps to solve the Floquet problem~\cite{leskesFloquetTheorySolidstate2010}. Given a time-periodic Hamiltonian $\ham(t)=\ham(t+{2\pi}/{\omega})$, its eigenstates have the form $\ket{\Psi(t)}=e^{-i\lambda t}\ket{\Phi(t)}$, where $\ket{\Phi(t)}=\ket{\Phi(t+{2\pi}/{\omega})}$ is periodic in time and $\lambda$ denotes the eigenenergy. 
The Schr\"{o}dinger equation $i\frac{\partial}{\partial t}\ket{\Psi(t)}=H(t)\ket{\Psi(t)}$ is then equivalent to
\begin{equation}
    \left(H(t)-i\frac{\partial}{\partial t}\right)\ket{\Phi(t)}=\lambda\ket{\Phi(t)}.
\end{equation}
Plugging in the Fourier expansions $\ket{\Phi(t)}=\sum_n\ket{\Phi_n} e^{-in\omega t}$, $H(t)=\sum_{n}H_n e^{-in\omega t}$, we obtain
\begin{align}
\left(\sum_{n}H_n(t)e^{-in\omega t}-i\frac{\partial}{\partial t}\right)\sum_p\ket{\Phi_p}e^{-im\omega t}=\lambda \sum_p\ket{\Phi_p}e^{-im\omega t}.
\end{align}
This can be expressed in  matrix form as $\ham_F\vec\Phi=\lambda\vec\Phi$,
\begin{equation} 
\begin{pmatrix}
 \ddots&\vdots &0&\vdots& \\ 
\cdots &H_0\!+\!\omega&H_{-1}&0&\cdots \\ 
 \cdots&H_1&H_0&H_{-1}& \\ 
 \cdots&0&H_1&H_0\!-\!\omega &\cdots \\ 
 &\vdots&\vdots&\vdots&\ddots
\end{pmatrix}\!\!
\begin{bmatrix}
\vdots\\
\Phi_{-1}\\
\Phi_0\\
\Phi_1\\
\vdots
\end{bmatrix}\!\!
=\!\lambda\!
 \begin{bmatrix}
\vdots\\
\Phi_{-1}\\
\Phi_0\\
\Phi_1\\
\vdots
\end{bmatrix},
\label{eq:FloquetMatrix}
\end{equation}
which  defines  the Floquet matrix $\ham_F$. 

Given the  Floquet matrix eigenenergies $\lambda^\mu$ and  eigenvectors $\vec\Phi^\mu=(\cdots,\Phi_{-1}^\mu,\Phi_0^\mu,\Phi_1^\mu,\cdots)^T$, the coherent state evolution of the system is a superposition of these eigenvectors
\begin{equation}
    \ket{\Psi(t)}=\sum_\mu c^\mu \ket{\Psi^\mu(t)}= \sum_\mu c^\mu e^{-i\lambda^\mu t}\sum_{n=-\infty}^{+\infty}e^{-in\omega t}\ket{\Phi_n^\mu}
\end{equation}
with coefficients $c^\mu$ determined by the initial condition
\begin{equation}
    \ket{\Psi(t=0)}=\sum_\mu c^\mu \sum_{n=-\infty}^{+\infty}\ket{\Phi_n^\mu}.
\end{equation}. 

\paragraph{a. Quasi energy.}
Given an eigenenergy $\lambda^\mu$, $\lambda^\mu+n\omega$ for any integer $n$ is also an eigenenergy since $\ket{\Psi^\mu(t)}=e^{-i\lambda^\mu t}\ket{\Phi^\mu(t)}=e^{-i\lambda^\mu t-in\omega t}(e^{in\omega t}\ket{\Phi^\mu(t)})$. To simplify the calculation, we usually limit the range of the eigenvalues within the first ``Brillouin zone''  $(-\omega/2,\omega/2]$. Note that the definition of ``Brillouin zone'' is flexible and in Ref.~\cite{wang_observation_2021} it is defined as $(0,\omega]$ to comply with the definition conventions of Mollow triplets. In this work, we define it as $(-\omega/2,\omega/2]$ to make it easier to analyze the particle-hole symmetry where two eigenvalues have symmetry-induced relation $\lambda^-=-\lambda^+$.

\paragraph{b. Number of nontrivial eigenvalues in an N-level quantum system.}
For an $N$-level quantum system, the Hamiltonian $\ham(t)$ is a $N\times N$ matrix, state vectors $\ket{\Psi(t)}$, $\ket{\Phi(t)}$ and their the Fourier components $\ket{\Phi_n}$,  are $N\times1$ vectors which can also be equivalently written as $\ket{\Psi(t)},\ket{\Phi(t)}$. 
Even in Floquet space,  there are $N$ non-trivial eigenvalues (in the first Brillouin zone) denoted by $\lambda^\mu$, while all the other eigenvalues are simple translations of these values with $\lambda^\mu+n\omega$.

\paragraph{c. Number of nontrivial eigenvalues in a qubit system.}
For qubit systems, the Hamiltonian $\ham(t)$ are $2\times2$ matrices, eigenstates $\ket{\Psi(t)}$, $\ket{\Phi(t)}$ and its the Fourier components $\ket{\Phi_n}$,  are $2\times1$ vectors which can also be equivalently written as $\ket{{\Psi(t)}},\ket{\Phi(t)}$. Due to the equidistant energy structures of a Floquet problem, there are two non-trivial eigenvalues denoted by $\lambda^\pm$ for a 2-level system and all the other eigenvalues are simple translation of these two values with $\lambda^\pm+n\omega$.

\paragraph{d. Truncation in numerical simulations.}
Since the Floquet matrix $\ham_F$ has infinite dimension, we truncate the matrix to $400\times400$ ($300\times 300$) blocks in the numerical simulation for the 2-level system (3-level system) in this work. In Figs.~\ref{CDT_evolution}(e-h) we compare the Floquet simulation to a simulation based on a Trotterized evolution, and show that the Floquet solution overlaps perfectly with the Trotter solution. Note that in the Trotterized evolution, we discretize the time to small steps $dt=0.0004\mu$s ($dt=0.008,0.00025\mu$s for Figs.~\ref{Spin1_simulation_Supp}(a,c)) and calculate the evolution by multiplying the time series of $e^{-iHdt}$ with the Hamiltonian in $\ham_I$.

Note that except for the plots of eigenstates in the Bloch sphere, Floquet eigenvalues and calculation of parameters $\pi_\mu$, all other simulations in this work is done by the exact simulation although the Floquet has the similar precision and efficiency.

\subsubsection{2. Generalized Rabi amplitudes}
Although the Rabi oscillation is typically defined for a(n effective) two-level system (e.g., measurement of $P_{\ket{0}}(t)$ for a qubit), a generalized Rabi oscillation $P_{\ket{k}}(t)$ can be defined for an $N$-level system where $k$ denotes the projection state. With the Floquet theory, such a Rabi oscillation can be calculated as
\begin{align}
        P_{\ket{k}}(t)&=\sum_n e^{-in\omega t}\sum_\mu c^\mu e^{-i\lambda^\mu t}\Phi^\mu_{n,k}\times
        \sum_p e^{im\omega t}\sum_\nu c^{\nu*}e^{i\lambda^\nu t}\Phi^{\nu *}_{p,k}\\
        &=\sum_{n,p}e^{i(p-n)\omega t}\left(\sum_\mu |c^\mu|^2\Phi_{p,k}^{\mu*}\Phi_{n,k}^{\mu}+\sum_{\mu,\nu\neq\mu}
        e^{-i(\lambda^\mu-\lambda^\nu)t}c^\mu c^{\nu*} \Phi_{n,k}^{\mu}\Phi_{p,k}^{\nu*}\right)\nonumber,
\end{align}
We can rewrite this expression as 
\begin{equation}
\label{Eq_P0_appendix}
  P_{\ket{k}}(t)  = \sum_{\mu,\nu,n} |a_{\mu,\nu,k}^{(n)}|\cos(\omega_{\mu,\nu,k}^{(n)} t+\phi_{\mu,\nu,n})
\end{equation}        
with frequency $\omega_{\mu,\nu,k}^{(n)}=n\omega+(\lambda^\mu-\lambda^\nu)$, and amplitudes $a_{\mu,\nu,k}^{(n)} = |a_{\mu,\nu,k}^{(n)}|\exp(i\phi_{\mu,\nu,k}^{(n)})=2\sum_p c^{\mu*}c^{\nu}\Phi_{p+n,k}^{\mu*}\Phi_{p,k}^{\nu}$.
Note that when making correspondence to the dynamical dipole matrix element of a probing operator ${V}$ in this work, the state $\ket{k}$ should be the eigenstates of probing operator ${V}$.

\subsection{B. Concatenated continuous driving}
Concatenated continuous driving (CCD) is a dynamical decoupling technique with modulated driving fields, and it has been explored  in the context of qubit coherence protection and noise decoupling~\cite{caiRobustDynamicalDecoupling2012,khanejaUltraBroadbandNMR2016,saikoSuppressionElectronSpin2018,cohenContinuousDynamicalDecoupling2017,farfurnikExperimentalRealizationTimedependent2017,rohrSynchronizingDynamicsSingle2014,laytonRabiResonanceSpin2014,saikoMultiphotonTransitionsRabi2015,teissierHybridContinuousDynamical2017,bertainaExperimentalProtectionQubit2020,caoProtectingQuantumSpin2020,wangCoherenceProtectionDecay2020,wang_observation_2021}. Although more complicated CCD schemes have been studied~\cite{caiRobustDynamicalDecoupling2012,khanejaUltraBroadbandNMR2016}, in this work we focus on the simplest one where only one modulation is applied to the main driving field. We can classify the CCD schemes into two types: amplitude-modulated CCD and phase-modulated CCD.

\textit{Amplitude-modulated CCD -} By applying an amplitude-modulated microwave with waveform $\omega_{\mu w}=\Omega\cos(\omega t)-2\epsilon_m\sin(\omega t)\cos(\omega_m t+\phi)$ to a qubit along the transverse $x$ axis $\omega_{\mu w}\sx$, the Hamiltonian in the lab frame can be written as
\begin{equation}
    \ham=\frac{\omega_0}{2}\sz+\left(\Omega\cos(\omega t)-2\epsilon_m\sin(\omega t)\cos(\omega_m t+\phi)\right)\sx
    \label{Hamiltonian}
\end{equation}
where $\omega_0$ is the level splitting of the qubit, $\Omega,\epsilon_m$ are the driving and modulation strengths. 
We assume $\Omega,\epsilon_m\ll\omega_0$ and $\delta=\omega-\omega_0\ll\omega_0$ (as typically the power of AC fields is limited) such that the rotating wave approximation (RWA) is valid. In the first rotating frame defined by a unitary transformation $U=\exp(-i\frac{\omega t}{2}\sz)$, and applying the RWA, the Hamiltonian becomes
\begin{equation}
    \ham_I=-\frac{\delta}{2}\sz+\frac{\Omega}{2}\sx+\epsilon_m\cos(\omega_mt+\phi)\sy.
    \label{HI_AmpMod}
\end{equation}

\textit{Phase-modulated CCD -} In the phase-modulated scheme, the driving waveform has a time-dependent phase and the Hamiltonian in the lab frame is
\begin{equation}
    \ham=\frac{\omega_0}{2}\sz+\Omega\cos\left(\omega t+\frac{2\epsilon_m}{\omega_m}\cos(\omega_m t+\phi)\right)\sx.
    \label{Hamiltonian_PhaseMod}
\end{equation}
The rotating frame  transformation to simplify the analysis is defined by $U=\exp(-i\int_0^t H_0(t^{'})dt^{'})$ with $\ham_0(t)=({\omega}/{2})\sz- ({\epsilon_m\omega_m}/{\omega_m})\sin(\omega_m t+\phi)\sz$ and then $U=\exp\left[-i\left(\frac{\omega t}{2}\sz+\epsilon_m \frac{\cos(\omega_m t+\phi)}{\omega_m}\sz \right)\right]$. Assuming that the RWA is valid, the Hamiltonian in the interaction picture is
\begin{equation}
        \ham_I=-\frac{\delta}{2}\sz+\frac{\Omega}{2}\sx+\epsilon_m\sin(\omega_mt+\phi)\sz.
    \label{HI_PhaseMod}
\end{equation}

Both amplitude-modulated CCD and phase-modulated CCD can implement similar Hamiltonian $\ham_I$ comprised of a static field and an oscillating field. However, the amplitude modulation is applied by varying the second driving amplitude which suffers more from power fluctuations, while the phase modulation is applied by varying the driving phase $(2\epsilon_m/\omega_m)\cos(\omega_mt+\phi)$ and is robust against power fluctuations (although noise associated with imperfect resolution and faulty electronic elements are still possible). The phase-modulated CCD usually has better performance such as longer coherence time \cite{farfurnikExperimentalRealizationTimedependent2017,cohenContinuousDynamicalDecoupling2017,wangCoherenceProtectionDecay2020}, and less power limitations which enable larger $\epsilon_m$. 

\subsection{C. Floquet solution to a driven qubit}
We consider the Hamiltonian of the phase-modulated concatenated continuous driving in the interaction picture $\ham_I=({\Omega}/{2})\sx+\epsilon_m\sin(\omega_mt+\phi)\sz$. Fourier decomposition of this Hamiltonian gives $\ham_{I,0} = ({\Omega}/{2})\sx,\ H_{I,\pm1} = \mp[{\epsilon_m e^{\mp i\phi}}/({2i})]\sz$. The Floquet matrix problem is
\begin{equation} 
\begin{pmatrix}
 \ddots&\vdots&\vdots&\vdots&\vdots&\vdots&\vdots& \\ 
\cdots &\omega_m &\frac{\Omega}{2} &-i\frac{\epsilon_m}{2}e^{i\phi} &0 &0&0&\cdots \\ 
 \cdots&\frac{\Omega}{2} &\omega_m &0 &i\frac{\epsilon_m}{2}e^{i\phi} &0 &0 &\cdots\\
 \cdots&i\frac{\epsilon_m}{2}e^{-i\phi}  &0& 0 &\frac{\Omega}{2}&-i\frac{\epsilon_m}{2}ie^{i\phi}&0&\cdots\\
 \cdots&0 &-i\frac{\epsilon_m}{2}e^{-i\phi} &\frac{\Omega}{2} &0 &0&i\frac{\epsilon_m}{2}e^{i\phi} &\cdots\\
 \cdots&0 &0 &i\frac{\epsilon_m}{2}e^{-i\phi} &0&-\omega_m &\frac{\Omega}{2} &\cdots\\
 \cdots&0 &0 &0 & -i\frac{\epsilon_m}{2}e^{-i\phi}  &\frac{\Omega}{2} &-\omega_m &\cdots\\
 &\vdots&\vdots&\vdots&\vdots&\vdots&\vdots&\ddots
\end{pmatrix}
\begin{pmatrix}
\vdots\\
\Phi_{-1,0}\\
\Phi_{-1,1}\\
\Phi_{0,0}\\
\Phi_{0,1}\\
\Phi_{1,0}\\
\Phi_{1,1}\\
\vdots
\end{pmatrix}
=\lambda
\begin{pmatrix}
\vdots\\
\Phi_{-1,0}\\
\Phi_{-1,1}\\
\Phi_{0,0}\\
\Phi_{0,1}\\
\Phi_{1,0}\\
\Phi_{1,1}\\
\vdots
\end{pmatrix}.
\end{equation}
Since we can always write the eigenstate $\ket{\Phi(t)},\ket{\Psi(t)}$ in the basis of ${V}=\sigma_j$ ($\ket{0_j}$, $\ket{1_j}$), the evolution of the system can be written in a more general form as
	\begin{equation}
	\ket{\Psi(t)}=c^+e^{-i\lambda^+t}\sum_n 
\begin{pmatrix}
    \Phi^+_{n,0_j}\\
    \Phi^+_{n,1_j}
    \end{pmatrix}
e^{-in\omega_m t}    +c^-e^{-i\lambda^-t}\sum_n 
\begin{pmatrix}
    \Phi^-_{n,0_j}\\
    \Phi^-_{n,1_j}
    \end{pmatrix}
e^{-in\omega_m t}= \sum_n 
\begin{pmatrix}
    c^+e^{-i\lambda^+t}\Phi^+_{n,0_j}+c^-e^{-i\lambda^-t}\Phi^-_{n,0_j}\\
    c^+e^{-i\lambda^+t}\Phi^+_{n,1_j}+c^-e^{-i\lambda^-t}\Phi^+_{n,1_j}
    \end{pmatrix}e^{-in\omega_m t}	
	\end{equation}    
where the coefficients $c^\pm$ are given by the initial condition $\ket{\Psi(0)}=\sum_\pm c^\pm\ket{\Phi^\pm(0)}$, and $0_j,1_j$ denotes the component on the basis $\ket{0_j},\ket{1_j}$ (for $\sigma_j=\sx,\sy,\sz$, the basis $\ket{0_j}=\ket{+},\ket{+i},\ket{0}$). 

For the Rabi measurements in this work, the qubit state is projected onto $|0_j\rangle$, the eigenstate of the probing operator ${V}\equiv\sigma_j$. 
The probability of being in the $|0_j\rangle$ state $P_{|0_j\rangle}(t)$ presents three classes of frequencies: $n\omega_m\pm(\lambda^+-\lambda^-)$, $n\omega_m$ where $n$ is the integer denoting the order of the bands:
\begin{align}
        P_{|0_j\rangle}(t)&=\sum_n e^{-in\omega_m t}(c^+e^{-i\lambda^+t}\Phi^+_{n,0_j}+c^-e^{-i\lambda^-t}\Phi^-_{n,0_j})\times
        \sum_p e^{im\omega_m t}(c^{+*}e^{i\lambda^+t}\Phi^{+*}_{p,0_j}+c^{-*}e^{i\lambda^-t}\Phi^{-*}_{p,0_j})\\
        &=\sum_{n,p}e^{i(p-n)\omega_m t}\left(|c^+|^2\Phi_{p,0_j}^{+*}\Phi_{n,0_j}^{+}+|c^-|^2\Phi_{p,0_j}^{-*}\Phi_{n,0_j}^{-}+
        e^{-i(\lambda^+-\lambda^-)t}c^+c^{-*} \Phi_{n,0_j}^{+}\Phi_{p,0_j}^{-*}
        +e^{i(\lambda^+-\lambda^-)t}c^{+*}c^{-}\Phi_{p,0_j}^{+*}\Phi_{n,0_j}^{-}\right)\nonumber
\end{align}
We can rewrite this expression as 
\begin{equation}
\label{Eq_P0_appendix}
  P_{|0_j\rangle}(t)  = \sum_{i,n} |a_{i}^{(n)}|\cos(\omega_{i}^{(n)} t+\phi_{i}^{(n)})
\end{equation}        
with $\omega_{i}^{(n)}=n\omega_m+i(\lambda^+-\lambda^-)$, where $i=-1,0,1$, $a_{\pm 1,n} = |a_{\pm 1,n}|\exp(i\phi_{\pm 1,n})=2\sum_p c^{\pm*}c^{\mp}\Phi_{p+n,0_j}^{\pm*}\Phi_{p,0_j}^{\mp}$, and $a_{0}^{(n)}=2\sum_{\pm}|c^\pm|^2\sum_p \Phi_{p+n,0_j}^{\pm*}\Phi_{p,0_j}^{\pm}$

\section{S2. Correspondence between Rabi amplitudes and susceptibility}
In Ref.~\cite{engelhardt_dynamical_2021}, the Floquet band is probed through the susceptibility $\chi$, as typically done in  light scattering experiments. Ref.~\cite{wang_observation_2021}, instead, probed the Floquet band  through the Rabi amplitudes of the coherent evolution of a driven qubit. Before proceeding to derive the selection rule in the Rabi measurement, we first establish the correspondence between the Rabi amplitudes and susceptibility $\chi$. 

\textit{Susceptibility $\chi$ - } In Ref.~\cite{engelhardt_dynamical_2021}, the Floquet band structure of a periodically driven qubit is probed through a bichromatic probe field ($\omega_{p,1}$ and $\omega_{p,2}=\omega_{p,1}+n\omega_m$) where the intensity change of the second frequency component $\Delta I(\omega_{p,2})=-i\chi_n(\omega_{p,1})\langle\hat{a}_{\omega_{p,2}}^\dagger\hat{a}_{\omega_{p,1}}\rangle+c.c.$ is used to identify the symmetry-protected selection rules. There, the susceptibility $\chi_n(\omega_{p,1})$ was defined as 
\begin{equation}
    \chi_n(\omega_{p,1})=i\lambda^2\sum_{\mu,\nu,m} \frac{V_{\nu,\mu}^{(-n-m)}V_{\mu,\nu}^{(m)}(p_\nu-p_\mu)}{\lambda^\mu-\lambda^\nu+m\omega_m-\omega_{p,1}-i\gamma_{\nu,\mu}^{(m)}}
    \label{susceptibility}
\end{equation}
where $\gamma$ is a phenomenological dephasing factor, and $\{p_\mu\}$ are associated with the long time stationary state $\rho(t)=\sum_\mu p_\mu \ket{\Phi^\mu(t)}\bra{\Phi^\mu(t)}$ is assumed in the derivation~\cite{engelhardt_dynamical_2021}.  
The symmetry-protected dark states (spDS) and dark bands (spDB) are induced by vanishing dynamical dipole matrix element $V_{\mu,\nu}^{(n)}=0$ and susceptibility $\chi_n(\omega_{p.1})=0$, respectively, and the symmetry-induced transparency (siT) is induced by the destructive interference of two nonzero dynamical dipole matrix elements $V_{\mu,\nu}^{(n)}\neq 0$.

\textit{$N$-level system - } In this work, we discuss the selection rule in the context of a coherent state evolution, and we find that the spDSs, spDBs, and siT also exist in the coherent state evolution under similar conditions. According to the Floquet calculation, the state evolution of a periodically-driven quantum system is $\ket{\Psi(t)}=\sum_\mu c^\mu e^{-i\lambda^\mu t} \ket{\Phi^\mu(t)}$ where $\ket{\Phi^\mu(t)}=\sum_{n=-\infty}^{+\infty} e^{-in\omega t}\ket{\Phi_n^\mu}$ and $c^\mu$ is determined by the initial condition $\ket{\Psi(0)}=\sum_\mu c^\mu \ket{\Phi(0)}$. Assuming the probing operator can be written as ${V}=\sum_{k} \mathcal{V}_k\ket{k}\bra{k}$, the dynamical dipole matrix element $V_{\mu,\nu}^{(n)}$ can be calculated as
\begin{align}
    V_{\mu,\nu}^{(n)}&=\frac{1}{T}\int_0^T \langle\Phi^\mu(t)|{V}|\Phi^\nu(t)\rangle e^{-in\omega t}dt, 
\end{align} 
Using 
$\ket{\Phi^{\mu}(t)}=\sum_n \ket{\Phi^\mu_n}e^{-in\omega t}$ and $V=\sum_k\mathcal V_k\ket k\!\bra k$, we have 
\begin{equation}
     V_{\mu,\nu}^{(n)}=\sum_k \mathcal V_k\sum_p\bra{\Phi_p^\mu}k\rangle\langle k\ket{\Phi_{p-n}^\nu}=\sum_{p,k}\mathcal{V}_k\Phi_{p,k}^{\mu*}\Phi_{p-n,k}^{\nu}.
\end{equation}
where $\Phi^{\mu*}_{p,k}=\langle \Phi_p^\mu\ket{k},\Phi_{p-n,k}^{\nu}=\langle k\ket{\Phi_{p-n}^\nu}$.
We define a ``weighted Rabi oscillation'' for the $N$-level system as  
\begin{equation}
    P(t)=\sum_k \frac{\mathcal{V}_k}{\mathcal{V}} P_{\ket{k}}(t)
\end{equation} where $\mathcal{V}=\sum_k |\mathcal{V}_k|$ is the normalization factor and $P_{\ket{k}}(t)=|\bra k\Psi(t)\rangle|^2$ is the measured population on $\ket{k}$. Then we have 
\begin{equation}
    \mathcal{V}P(t)=\langle V\rangle=\bra{\Psi(t)}V\ket{\Psi(t)}
\end{equation}
with $\ket{\Psi(t)}=\sum_\mu c^\mu \ket{\Phi^\mu(t)}e^{-i\lambda^\mu t}=\sum_{\mu,n}c^\mu\ket{\Phi_n^\mu}e^{-in\omega t}e^{-i\lambda^\mu t}$, we obtain 
\begin{equation}
\langle V\rangle = \sum_{\mu,\nu,n}c^{\mu*} c^{\nu} e^{i(\lambda^\mu-\lambda^\nu)t}e^{in\omega t}V_{\mu,\nu}^{(n)}
\end{equation}        

This weighted Rabi oscillation can then be decomposed to a series of frequency components as
\begin{equation}
\label{Eq_P0_appendix}
  P(t) =\frac{\langle V\rangle}{\mathcal{V}}  = \sum_{\mu,\nu,n} |a_{\mu,\nu}^{(n)}|\cos(\omega_{\mu,\nu}^{(n)} t+\phi_{\mu,\nu}^{(n)}),
\end{equation}        
where frequencies $\omega_{\mu,\nu}^{(n)}=n\omega+(\lambda^\mu-\lambda^\nu)$, amplitudes $a_{\mu,\nu}^{(n)} = |a_{\mu,\nu}^{(n)}|\exp(i\phi_{\mu,\nu}^{(n)})=2\sum_{p,k} c^{\mu*}c^{\nu}(\mathcal{V}_k/\mathcal{V}) \Phi_{p+n,k}^{\mu*}\Phi_{p,k}^{\nu}$. When $\mu=\nu$, the observed center bands are coherent interference of all centerbands resulting in amplitudes $a_{0}^{(n)} = |a_{0}^{(n)}|\exp(i\phi_{0}^{(n)})=2\sum_{\mu,p,k} |c^{\mu}|^2 (\mathcal{V}_k/\mathcal{V}) \Phi_{p+n,k}^{\mu*}\Phi_{p,k}^{\mu}$. Note that interference of sidebands can happen at energy degeneracy points $\lambda^\mu=\lambda^\nu$. For $N$-level systems, the frequency modes of the Rabi oscillation $P(t)$ are comprised of a series of equidistant manifolds, and one centerband $n\omega$ and $N(N-1)$ sidebands $n\omega+(\lambda^\mu-\lambda^{\nu})$ present within each manifold.

As a result, we obtain the correspondence between the weighted Rabi amplitudes and the dynamical dipole matrix element
\begin{align}
    a_{\mu,\nu\neq\mu}^{(n)}&=2c^{\mu *}c^\nu \frac{V_{\mu,\nu}^{(n)}}{\mathcal{V}},\\
    a_{0}^{(n)}&=2\sum_{\mu} |c^\mu|^2 \frac{V_{\mu,\mu}^{(n)}}{\mathcal{V}}.
\end{align}
As derived above, the off-diagonal terms of the dynamical dipole transition element $2c^{\mu *}c^\nu V_{\mu,\nu\neq\mu}^{(n)}/\mathcal{V}$ can be exactly mapped to the sidebands amplitudes $a_{\mu,\nu\neq\mu}^{(n)}$, while the weighted summation of the diagonal terms of the dynamical dipole transition element $2\sum_{\mu} |c^\mu|^2 V_{\mu,\mu}^{(n)}/\mathcal{V}$ can be mapped to the centerbands amplitudes $a_{0}^{(n)}$.

\textit{Quantum mode control - } Since we have coherent phase control over the applied microwave, we can selectively observe each individual centerband and sideband by tuning the coefficients $c^\mu$ through the initial condition (\textit{quantum mode control}) as discussed in both the main text and Ref.~\cite{wang_observation_2021}. 
As a result, all elements of the dynamical dipole matrix $V_{\mu,\nu}^{(n)}$ can be individually probed through  \textit{quantum mode control}. 
We clarify that  symmetry-protected selection rules discussed in this work determine the selection rule of $V_{\mu,\nu}^{(n)}$, while \textit{quantum mode control} determines via $c^\mu$ which  mode amplitudes are observed,   by tuning the initial state with respect to the Floquet eigenstates.

\textit{2-level system - } According to  Floquet theory, the state evolution of a driven qubit is $\ket{\Psi(t)}=\sum_\pm c^\pm e^{-i\lambda^\pm t} \ket{\Phi^\pm(t)}$ where $\ket{\Phi^\pm(t)}=\sum_{n=-\infty}^{+\infty} e^{-in\omega_mt}\ket{\Phi_n^\pm}$. For the two-level qubit, the probing operators $\sx,\sz,\sy$ can be written as $\sx=\ket{+}\bra{+}-\ket{-}\bra{-},\sy=\ket{+i}\bra{+i}-\ket{-i}\bra{-i},\sx=\ket{0}\bra{0}-\ket{1}\bra{1}$, thus the dynamical dipole matrix element $V_{\mu,\nu}^{(n)}$ can be simply calculated as 
\begin{align}
    V_{\mu,\nu}^{(n)}&=\frac{1}{T}\int_0^T \langle\Phi^\mu(t)|{V}|\Phi^\nu(t)\rangle e^{-in\omega t}dt=\sum_p\left(\Phi_{p,0_j}^{\mu*}\Phi_{p-n,0_j}^{\nu}-\Phi_{p,1_j}^{\mu*}\Phi_{p-n,1_j}^{\nu}\right)
\end{align}
where $\mu,\nu\in\{+,-\}$ corresponding to the two Floquet eigenstates, and $0_j,1_j$ denote the projection on the two eigenstates of the probing operator ${V}=\sigma_j$. Since $P_{\ket{0_j}}(t)+P_{\ket{1_j}}(t)=1$, the measurement of $(1/2)\left[P_{\ket{0_j}}(t)-P_{\ket{1_j}}(t)\right]=P_{\ket{0_j}}(t)-1/2$, we can measure the typical qubit Rabi oscillation $P_{\ket{0_j}}$ instead of the equivalent population difference between $\ket{0_j}$ and $\ket{1_j}$ (except for the zero frequency component where the equivalence is no longer satisfied). 
The Rabi amplitudes $P_{\ket{0_j}}$ in Eq.~\eqref{Eq_P0_appendix} can be expressed in terms of the dynamical dipole matrix elements as listed in Table 1 of the main text.

\newpage
\section{S3. Observation of symmetries in a two-level system}
\label{Sec:Symmetries}
We apply the phase-modulated CCD technique to engineer a Hamiltonian in the interaction picture
\begin{equation}
        \ham_I=\frac{\Omega}{2}\sx+\epsilon_m\sin(\omega_mt+\phi)\sz.
    \label{HI_PhaseMod_resonance}
\end{equation}
The time period of such a Hamiltonian is $T=2\pi/\omega_m$.

\begin{figure}[b]
\includegraphics[width=\textwidth]{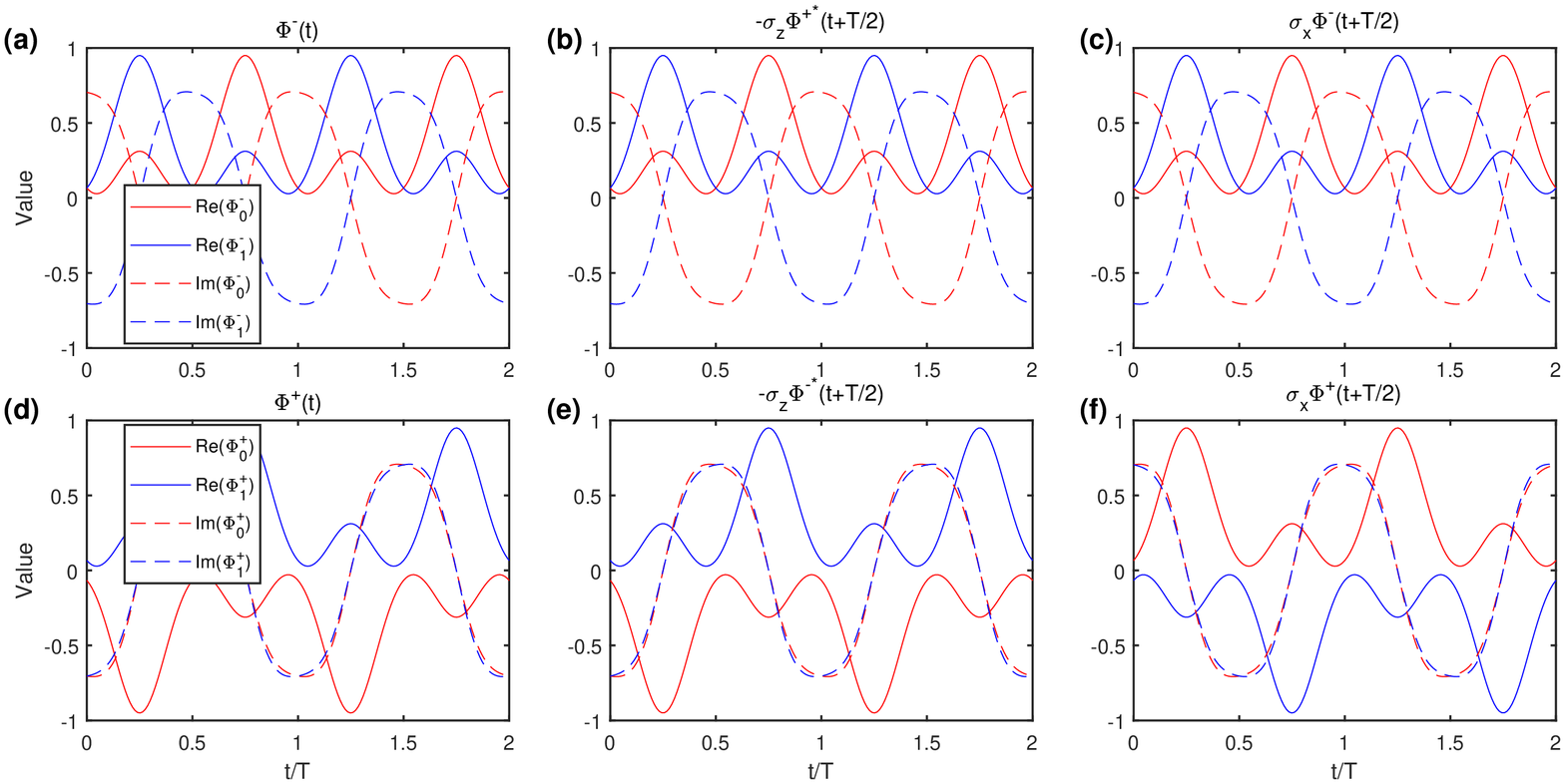}
\caption{\label{Eigenstate_Symmetry} Symmetry of Floquet eigenstates $\ket{\Phi^\pm(t)}$. Floquet simulation is applied to calculate the time-dependent Floquet eigenstates $\ket{\Phi^\pm(t)}$ under the condition $2\epsilon_m/\omega_m=2.75$, $\phi=0$, and $\omega_m=\Omega=(2\pi)3\text{MHz}$. (a) $\Phi^-(t)$ plotted in the first two time periods. Since $\Phi^-(t)$ is a $2\times1$ normalized complex vector, we plot the real (solid line) and imaginary (dashed lines) parts of each component  (red and blue colors). The basis used here is the $\sz$ basis. Same rules apply for the other plots. (b) $-\sz\Phi^{+*}(t+T/2)$. (c) $\sx\Phi^-(t+T/2)$.  (d) $\Phi^+(t)$. (e) $-\sz\Phi^{-*}(t+T/2)$. (f) $\sx\Phi^+(t+T/2)$. }
\end{figure}

\begin{figure}[h]
\includegraphics[width=\textwidth]{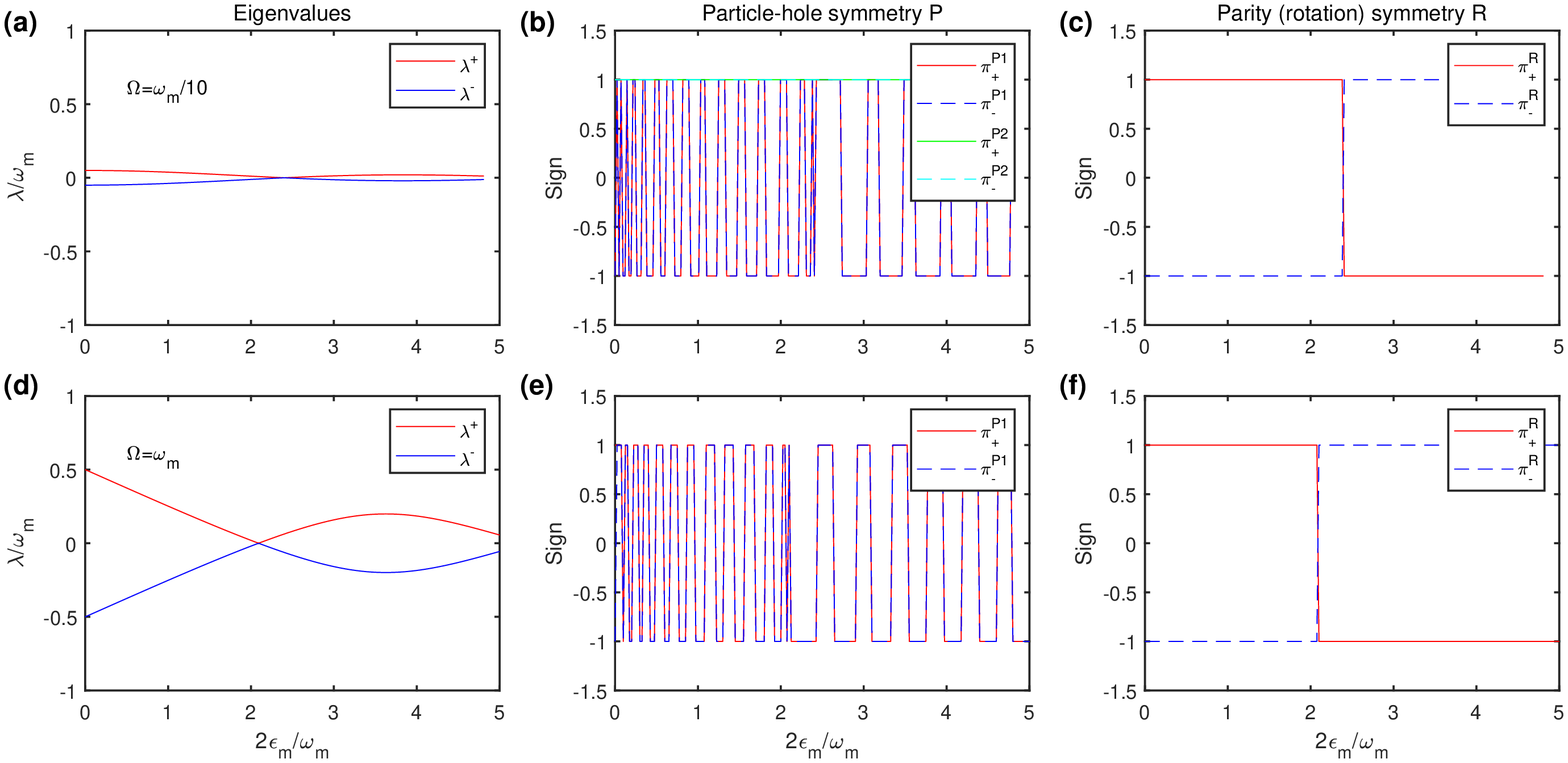}
\caption{\label{Eigenstate_Symmetry_2} Floquet eigenenergies and $\pi_\mu$. (a) Floquet eigenenergies $\lambda^\pm$ under the condition $\phi=0$, $\omega_m=10\Omega=(2\pi)15\text{MHz}$. Note that $\lambda^\pm$ is defined in the range $(-\omega_m/2,\omega_m/2]$ and $\lambda^-\leq\lambda^+$. (b) The values of $\pi_\pm^{(P_1)}$, $\pi_\pm^{(P_2)}$ corresponding to the two particle-hole symmetries $\hat{P}_1=\sz$, $\hat{P}_2=I$ under the condition in (a). (c) The values of  $\pi_\pm^{(R)}$ corresponding to the parity symmetry $\hat{R}=\sx$ under the condition in (a). (d) Floquet eigenenergies $\lambda^\pm$ under the condition $\phi=0$, $\omega_m=\Omega=(2\pi)3\text{MHz}$. (e) The values of $\pi_\pm^{(P_1)}$ corresponding to the particle-hole symmetry $\hat{P}_1=\sz$ under the condition in (d). (f) The values of  $\pi_\pm^{(R)}$ corresponding to the parity symmetry $\hat{R}=\sx$ under the condition in (d).}
\end{figure}

\begin{figure}[h]
\includegraphics[width=\textwidth]{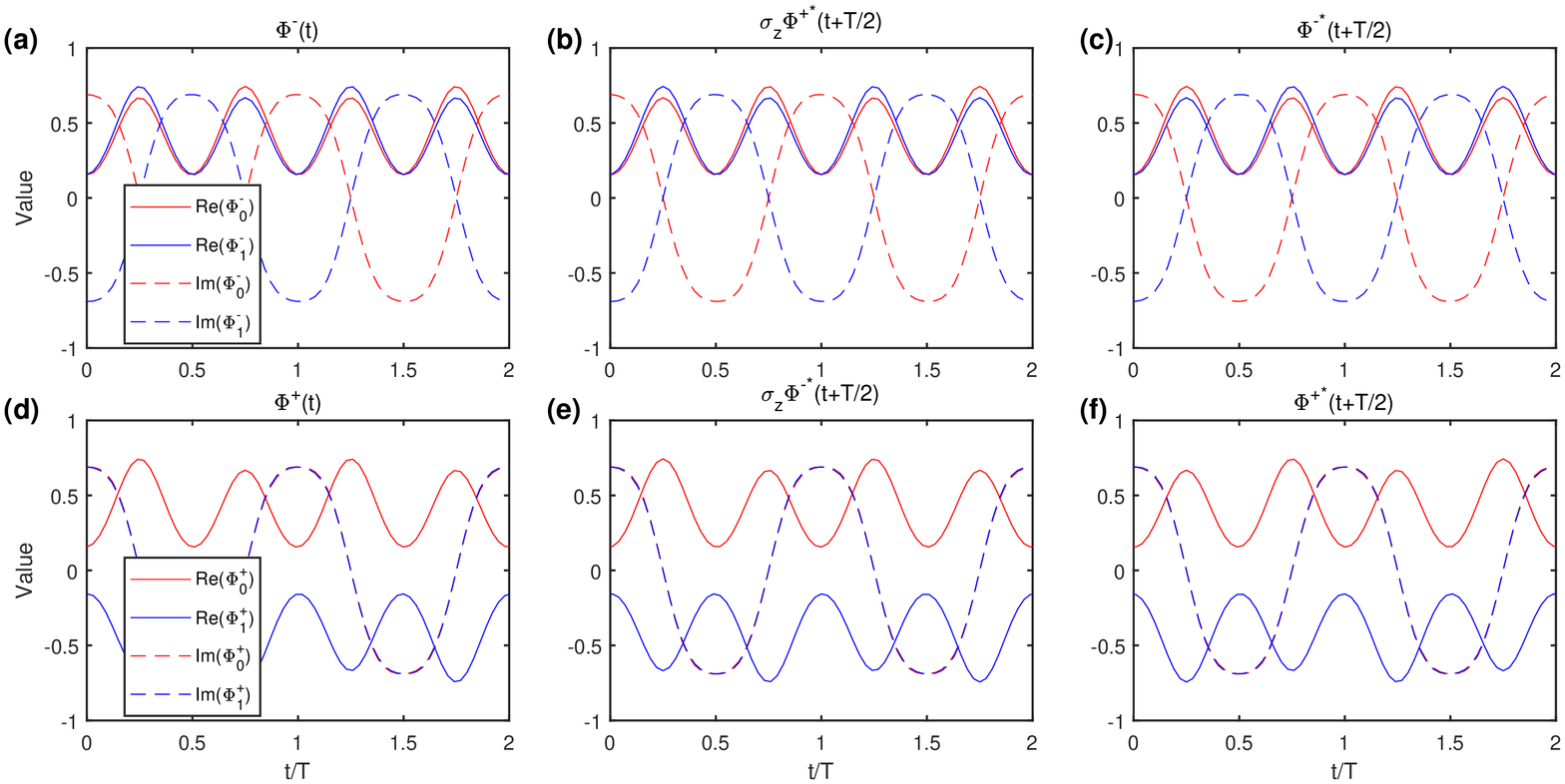}
\caption{\label{Eigenstate_Symmetry_twoPH} Symmetry of Floquet eigenstates $\ket{\Phi^\pm(t)}$. Floquet simulation is applied to calculate the time-dependent Floquet eigenstates $\ket{\Phi^\pm(t)}$ under the condition $2\epsilon_m/\omega_m=2.6934$, $\phi=0$, and $\omega_m=10\Omega=(2\pi)15\text{MHz}$. (a) $\Phi^-(t)$. (b) $\sz\Phi^{+*}(t+T/2)$. (c) $\Phi^{-*}(t+T/2)$.  (d) $\Phi^+(t)$. (e) $\sz\Phi^{-*}(t+T/2)$. (f) $\Phi^{+*}(t+T/2)$. }
\end{figure}

\subsection{A. Parity (2-fold rotation) symmetry $\sx$}
The first symmetry associated with the Hamiltonian $\ham_I$ in Eq.~\eqref{HI_PhaseMod_resonance} is a 2-fold rotation or, equivalently, a parity symmetry given by the operator $\hat{R}=\sx$, which simply gives $\hat{R}\ham_I(t+T/2)\hat{R^\dagger}=\ham_I(t)$. Such a symmetry gives rise to a relation $\ket{\Phi^\mu(t)}=\pi_\mu^{(R)}\hat{R}\ket{\Phi^\mu(t+T/2)}=e^{i\pi m_\mu}\hat{R}\ket{\Phi^\mu(t+T/2)}$ with $\mu\in\{-,+\}$ and $m_\mu\in\{0,1\}$~\cite{engelhardt_dynamical_2021}. In Fig.~\ref{Eigenstate_Symmetry} we numerically calculate and plot the Floquet eigenstates $\ket{\Phi^\pm(t)}$ under the resonance condition $\Omega=\omega_m=(2\pi)3\text{MHz}$, for a modulation strength $2\epsilon_m/\omega_m=2.75$ and phase  $\phi=0$. 
Under such a condition, $\pi_\pm^{(R)}=\pm 1$ as validated in the comparison between Figs.~\ref{Eigenstate_Symmetry}(a,d) and Figs.~\ref{Eigenstate_Symmetry}(c,f). 

The values of $\pi_\mu^{(R)}$ have dependence on the modulation parameters we choose (and also how we define the bands as discussed in Ref.~\cite{wang_observation_2021}). 
In Figs.~\ref{Eigenstate_Symmetry_2}(c,f) we plot the two values of $\pi_\pm^{(R)}$ as a dependence of $2\epsilon_m/\omega_m$ which switch with each other when the eigenvalues cross the degeneracy point (under the definition that $\lambda^-<\lambda^+$ and $\lambda^\pm\in(-\omega_m/2,\omega_m/2]$). Due to the parity symmetry (2-fold rotational symmetry), the relation $\pi_+^{(R)}\pi_-^{(R)}=-1$ is satisfied, which is used in the following selection rule derivations.

Now we start to derive the selection rule by evaluating the value of the dynamical dipole matrix element. 
The dynamical dipole matrix element is
\begin{align}
    V_{\mu,\nu}^{(n)}=\frac{1}{T}\int_0^T \langle\Phi^\mu(t)|{V}|\Phi^\nu(t)\rangle e^{-in\omega_mt}dt=\frac{1}{T}\int_0^{T/2} \langle\Phi^\mu(t)|{V}|\Phi^\nu(t)\rangle e^{-in\omega_mt}dt\times\left[1+\pi_\mu^{(R)}\pi_\nu^{(R)*}e^{-i\pi n}\alpha_V\right]\nonumber
\end{align}
where the value of $\alpha_V$ is given by the relation $\hat{R}{V}\hat{R}^\dagger=\alpha_V{V}$.
When the observation operator ${V}=\sy,\sz$ such that $\alpha_V=-1$, Mollow centerbands with odd $(n)$ orders are visible while the even orders are vanishing. For the sidebands, the even orders are visible while the odd orders are vanishing. When the observation operator ${V}=\sx$ such that $\alpha_V=1$, Mollow centerbands with even $(n)$ orders are visible while the odd orders are vanishing. For the sidebands, the odd orders are visible due to $\pi_+^{(R)}\pi_-^{(R)}=-1$ while the even orders are vanishing. These predictions are observed in both experiments in the main text (complete data also shown in Fig.~\ref{Resonance_data_Supp}) and simulations in Fig.~\ref{Resonance_Simulation}.

\begin{figure}[t]
\includegraphics[width=0.95\textwidth]{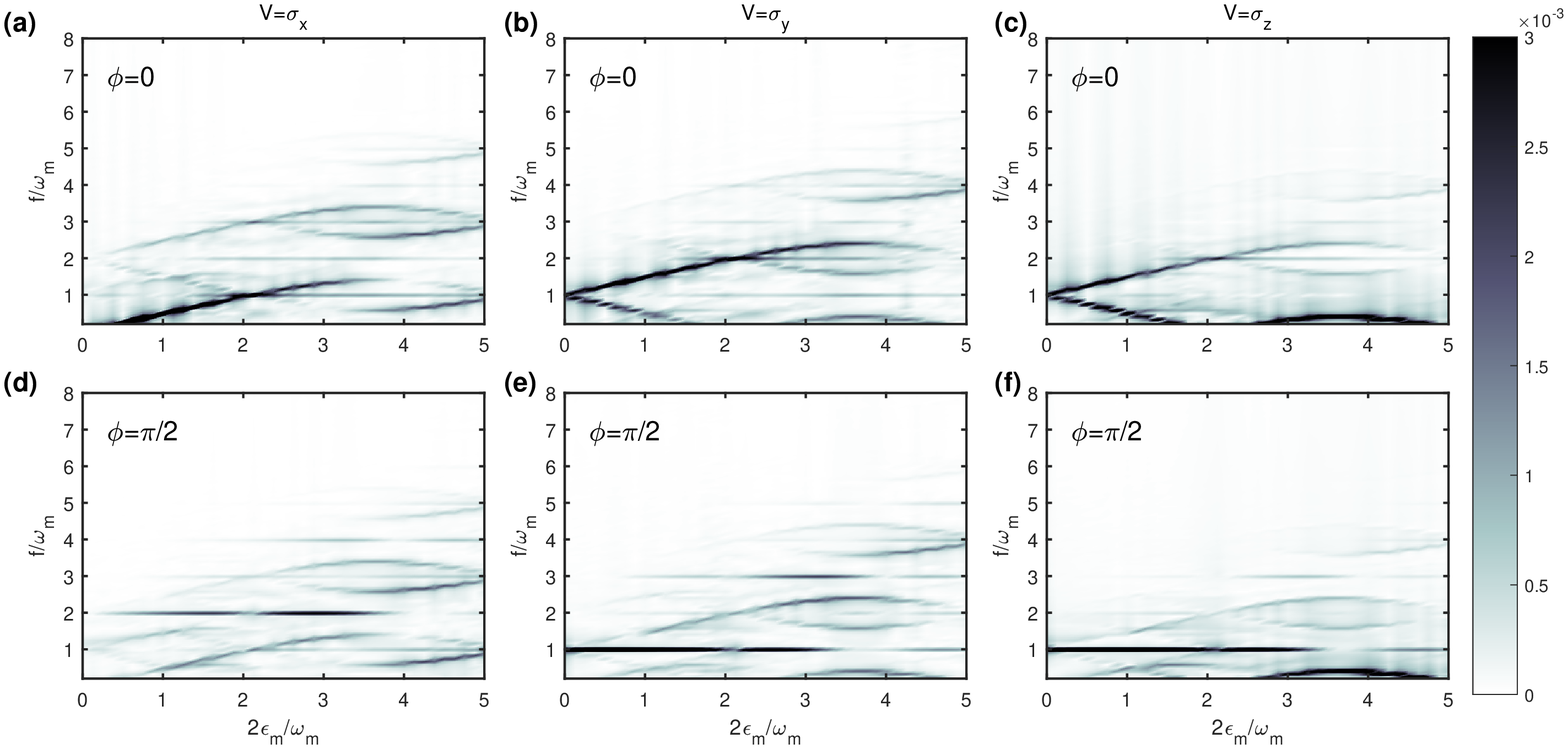}
\caption{\label{Resonance_data_Supp} Experimental observation of spDBs, spDSs, and destructive interference between centerbands. The Hamiltonian $\ham_I$ in Eq.~\eqref{HI_PhaseMod_resonance} is engineered by the phase-modulated CCD technique with parameters $\Omega=\omega_m=(2\pi)3\text{MHz}$, $\phi=0,\pi/2$ corresponding to (a-c,b-d). Rabi oscillations of an initial state $\ket{0}$ from $t=0$ to $t=4\mu$s are measured with 401 sampling points and their Fourier spectrum are plotted as intensity map. (a,d) Fourier spectrum of Rabi measurement on $\ket{+}$ (${V}=\sx$) under different modulation strength $2\epsilon_m/\omega_m$. (b,e) Fourier spectrum of Rabi measurement on $\ket{+i}$ (${V}=\sy$) under different modulation strength $2\epsilon_m/\omega_m$. (c,f) Fourier spectrum of Rabi measurement on $\ket{0}$ (${V}=\sz$) under different modulation strength $2\epsilon_m/\omega_m$.}
\end{figure}
\begin{figure}[htbp]
\includegraphics[width=\textwidth]{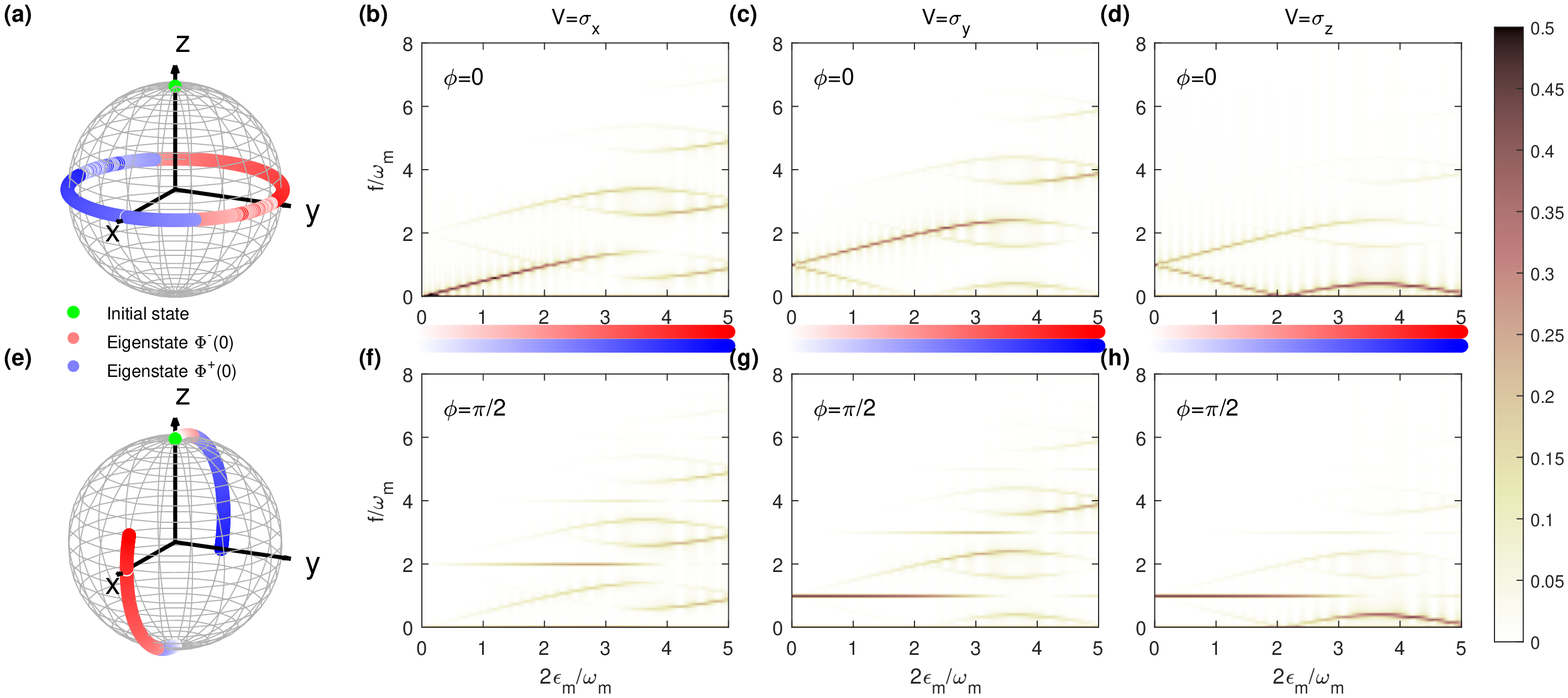}
\caption{\label{Resonance_Simulation} Simulation of spDBs, spDSs, and destructive interference between centerbands under the Hamiltonian $\ham_I$ in Eq.~\eqref{HI_PhaseMod_resonance}. (a,e) Floquet eigenstates $\ket{\Phi^\pm(0)}$ visualized on a Bloch sphere. Floquet eigenstates are calculated with the Hamiltonian in Eq.~\eqref{HI_PhaseMod_resonance}. Modulation strength $2\epsilon_m/\omega_m$ is swept from $0$ to $5$ with darker colors, and other parameters are $\Omega=\omega_m=(2\pi)3\text{MHz}$, $\phi=0,\pi/2$ corresponding to (a-d), (e-h) respectively. (b,f) Fourier spectrum of Rabi measurement on $\ket{+}$ (${V}=\sx$) under different modulation strength $2\epsilon_m/\omega_m$. Rabi oscillations of an initial state $\ket{0}$ from $t=0$ to $t=4\mu$s are simulated and their Fourier spectrum are plotted as intensity map. (c,g) Fourier spectrum of Rabi measurement on $\ket{+i}$ (${V}=\sy$) under different modulation strength $2\epsilon_m/\omega_m$. Other parameters are the same as in (b,f). (d,h) Fourier spectrum of Rabi measurement on $\ket{0}$ (${V}=\sz$) under different modulation strength $2\epsilon_m/\omega_m$. Other parameters are the same as in (b,f).}
\end{figure}

\subsection{B. Particle-hole symmetry $\sz$}
The second symmetry is a particle-hole symmetry $\hat{P_1}=\sz$, which gives $\hat{P_1}\ham_I(t+T/2)\hat{P_1}^\dagger=-\ham_I(t)$. Given an eigenstate $\ket{\Phi^\mu(t)}$ corresponding to an eigenenergy $\lambda^\mu$, its counterpart $\hat{P_1}\ket{\Phi^{\mu *}(t+T/2)}$ is also an eigenstate of the system with eigenenergy $-\lambda^\mu$. For our qubit system with two nontrivial eigensolutions (two eigenvalues are defined within the first ``Brillouin zone''  $(-\omega_m/2,\omega_m/2]$ in this work), such a relation yields $\ket{\Phi^{-\mu}(t)}=\pi_\mu^{(P_1)}\hat{P_1}\ket{\Phi^{\mu *}(t+T/2)}$ where  $\pi_\mu^{(P_1)}$ is gauge-dependent~\cite{engelhardt_dynamical_2021}. 

We have $\pi_\mu^{(P_1)}\in\{1,-1\}$ as validated by the Floquet calculation in the comparison between Figs.~\ref{Eigenstate_Symmetry}(a,d) and Figs.~\ref{Eigenstate_Symmetry}(b,e). From the equality $\ket{\Phi^\mu(t)}=\pi_{\mu^\prime}^{(P_1)}\hat{P_1}\ket{\Phi^{\mu^\prime * }(t+T/2)}=\pi_{\mu^\prime}^{(P_1)}\hat{P_1}\pi_{\mu}^{(P_1)*}\hat{P_1}^*\ket{\Phi^{\mu}(t+T)}=\pi_\mu^{(P_1)}\pi_{\mu^\prime}^{(P_1)*}\ket{\Phi^\mu(t)}$, we obtain that $\pi_\mu^{(P_1)}\pi_{\mu^\prime}^{(P_1)*}=1$, although the specific values of $\pi_\mu^{(P_1)}$ here depends on the parameter $2\epsilon_m/\omega_m$. In Figs.~\ref{Eigenstate_Symmetry_2}(b,e), we plot the values of $\pi_\mu^{(P_1)}$ for conditions $\Omega=\omega_m/10$, $\Omega=\omega_m$ as a dependence of modulation strength $2\epsilon_m/\omega_m$, where $\pi_\mu^{(P_1)}\pi_{\mu^\prime}^{(P_1)}=1$ is always satisfied although the value of $\pi_\mu^{(P_1)}$ depends on the modulation parameter.

The dynamical dipole matrix element can be calculated as 
\begin{align}
\label{Eq:derive_P1}
        V_{\mu,\nu}^{(n)}=&\frac{1}{T}\int_0^T \langle\Phi^\mu(t)|{V}|\Phi^\nu(t)\rangle e^{-in\omega_mt}dt=\frac{1}{T}\int_0^{T} \langle\Phi^{\mu^\prime *}(t+T/2)|\pi_{\mu^\prime}^{(P_1) *}\hat{P_1}^\dagger{V}\hat{P_1}\pi_{\nu^\prime}^{(P_1)}|\Phi^{\nu^\prime *}(t+T/2)\rangle  e^{-in\omega_mt}dt 
        \nonumber\\=&\alpha_V^{(P_1)}e^{i\pi n}V_{\mu^\prime,\nu^\prime}^{(-n) *}=\alpha_V^{(P_1)}e^{i\pi n}V_{\nu^\prime,\mu^\prime}^{(n)}
\end{align}
where $\alpha_V^{(P_1)}$ is given by the relation $\hat{P_1}^\dagger{V}\hat{P_1}=\alpha_V^{(P_1)}{V}^*$. Since the particle-hole symmetry operator $\hat{P_1}=\sz$ maps the Floquet eigenstate to its counterpart rather than itself as the parity symmetry, we have to consider the following two situations.

(1) When $\mu\neq\nu$, this gives the selection rule for the Mollow sidebands with $(\mu,\nu)=(\nu^\prime,\mu^\prime)$ such that $V_{\mu,\nu}^{(n)}=\alpha_V^{(P_1)}e^{i\pi n}V_{\mu,\nu}^{(n)}$. When the observation operator ${V}=\sz$, $\alpha_V^{(P_1)}=1$, Mollow sidebands are only visible with even order $(n)$. When the observation operator ${V}=\sx$, $\alpha_V^{(P_1)}=-1$, Mollow sidebands are only visible with odd order. This is consistent with the prediction from the parity symmetry. 

(2) When $\mu=\nu$, this gives rise to the vanishing centerbands when $|c^+|^2=|c^-|^2=1/2$ (the initial state is an equal superposition of two Floquet eigenstates) and $\alpha_V^{(P_1)}e^{i\pi n}=-1$. 
When the observation operator ${V}=\sz$, $\alpha_V^{(P_1)}=1$, Mollow centerbands are only visible with even order $(n)$ since $V_{+,+}^{(n)}=-V_{-,-}^{(n)}$ when $\alpha_V e^{-i\pi n}=-1$ for odd $n$. 
When the observation operator ${V}=\sx$, $\alpha_V^{(P_1)}=-1$, Mollow centerbands are only visible with odd order. 
However, due to the parity symmetry that makes these visible bands vanish, no centerband is visible for both cases when taking into account both the particle-hole symmetry and the parity symmetry. 
Such a prediction is observed in the comparison between $\phi=0$ and $\phi=\pi/2$ in both experiment (Fig.~\ref{Resonance_data_Supp}) and simulation (Fig.~\ref{Resonance_Simulation}) where centerbands are only visible for $\phi=\pi/2$ when $|c^\pm|^2\neq 1/2$. 
In Figs.~\ref{Resonance_Simulation}(a,e), we plot  the Floquet eigenstates on the Bloch sphere. Under the modulation phase $\phi=0$, the Floquet eigenstate is always in $x-y$ plane such that $|c^\pm|^2=1/2$ is always satisfied. Under the modulation phase $\phi=0$, the Floquet eigenstate is in $x-z$ plane such that $|c^\pm|^2\neq 1/2$ except for the modulation strength $2\epsilon_m/\omega_m\approx 4$ where  $|c^\pm|^2=1/2$ is accidentally satisfied. 

Note that $\hat{P}\hat{P}^*=I$ is required for particle-hole symmetry, thus when considering the amplitude-modulated CCD scheme, where the periodically driving fields are along $y$ direction, a frame transformation to align the driving fields to the $z$ direction is needed before applying  particle-hole symmetry analysis.

\clearpage
\subsection{C. Symmetry Breaking}
To demonstrate that indeed the selection rules arise from the symmetries of the  Hamiltonian, we spoil such symmetries by introducing additional terms.
The complete experimental data under the resonance condition $\Omega=\omega_m=(2\pi)3$MHz with engineered symmetry breaking Hamiltonian plotted in Fig.~\ref{ResonanceSB_data_Supp}, where the appearance of more forbidden bands clearly signals the symmetry breaking. Corresponding simulation is plotted in Fig.~\ref{ResonanceSB_Simulation}. See also section~\ref{sec:experiment} for experimental limitations that might induce weak breaking of the symmetries.

\begin{figure}[h]
\includegraphics[width=0.95\textwidth]{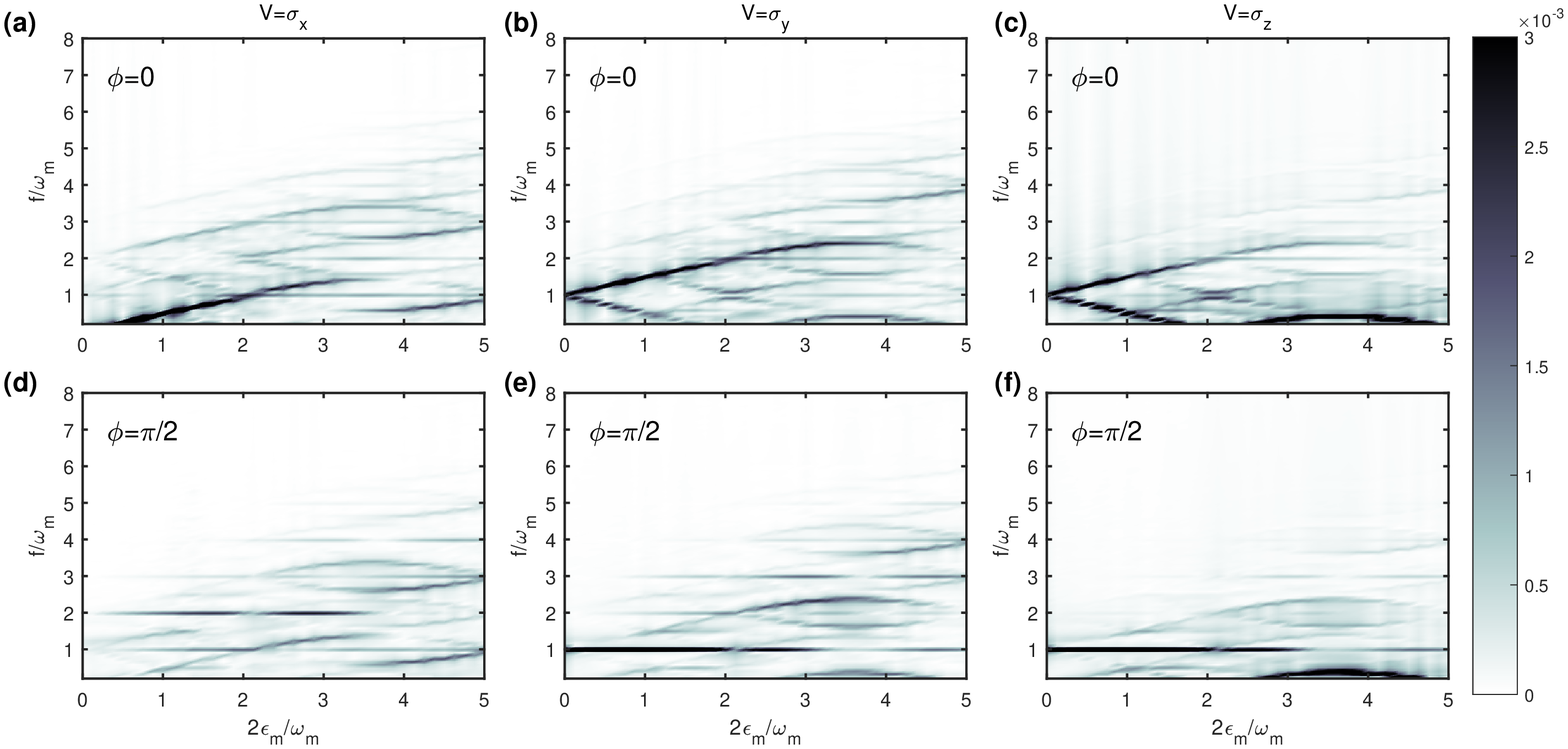}
\caption{\label{ResonanceSB_data_Supp} Experimental observation of symmetry breaking. Same experiment conditions are used as in Fig.~\ref{Resonance_data_Supp} except for the engineered Hamiltonian $\ham=(\Omega/2)\sx+\epsilon_m\sin(\omega_mt+\phi)+0.2\epsilon_m\sin(2\omega_mt+\phi)$ where the third term breaks both the parity symmetry and particle-hole symmetry.}
\end{figure}
\begin{figure}[h]
\includegraphics[width=0.95\textwidth]{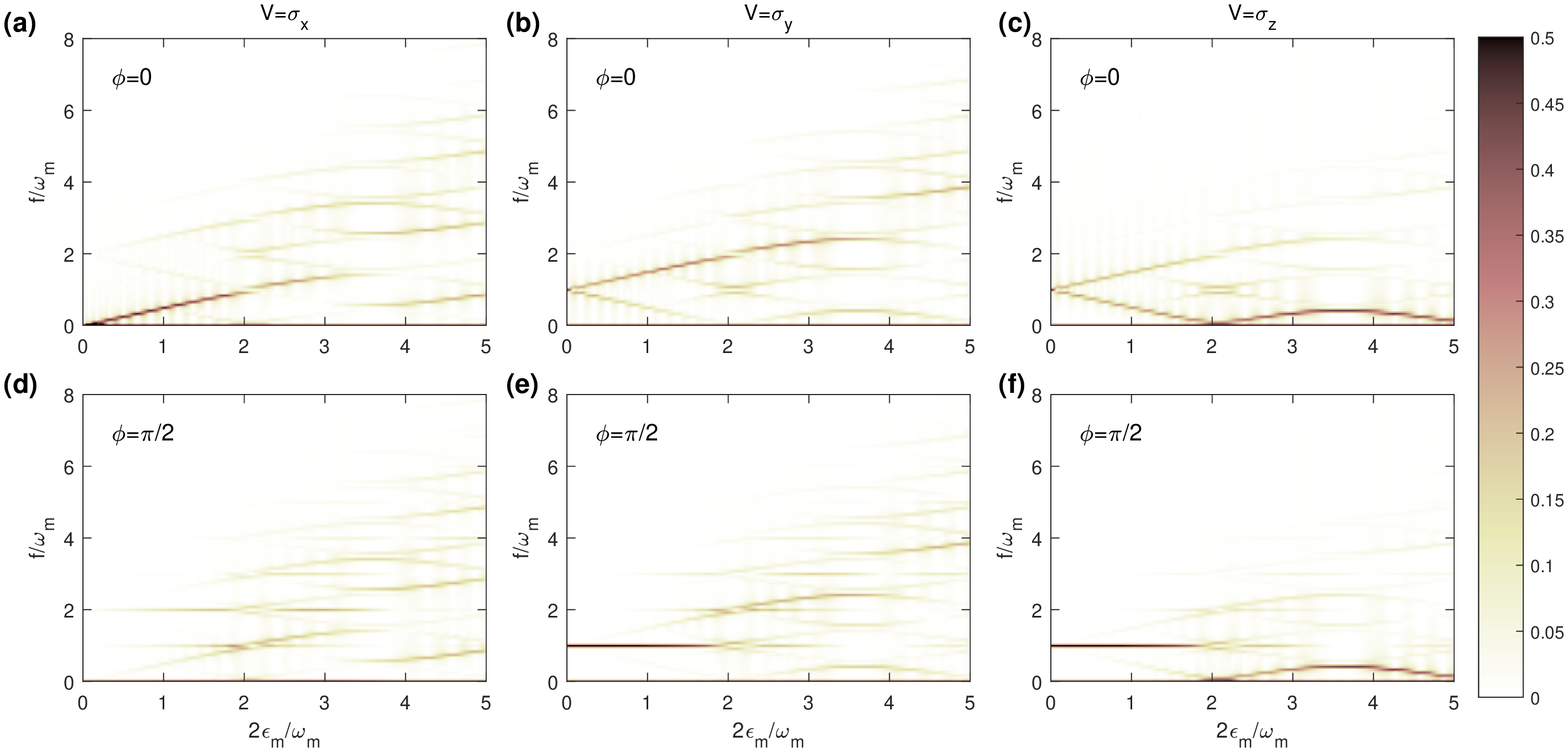}
\caption{\label{ResonanceSB_Simulation} Simulation of symmetry breaking. Same parameters are used as in Fig.~\ref{Resonance_Simulation} except for the Hamiltonian $\ham=(\Omega/2)\sx+\epsilon_m\sin(\omega_mt+\phi)\sz+0.2\epsilon_m\sin(2\omega_mt+\phi)\sz$ where the third term breaks both the parity symmetry and particle-hole symmetry.}
\end{figure}

\clearpage
\section{S4. Symmetry-induced transparency (siT or CDT)}\label{CDT studies}
\begin{figure}[h]
\includegraphics[width=0.7\textwidth]{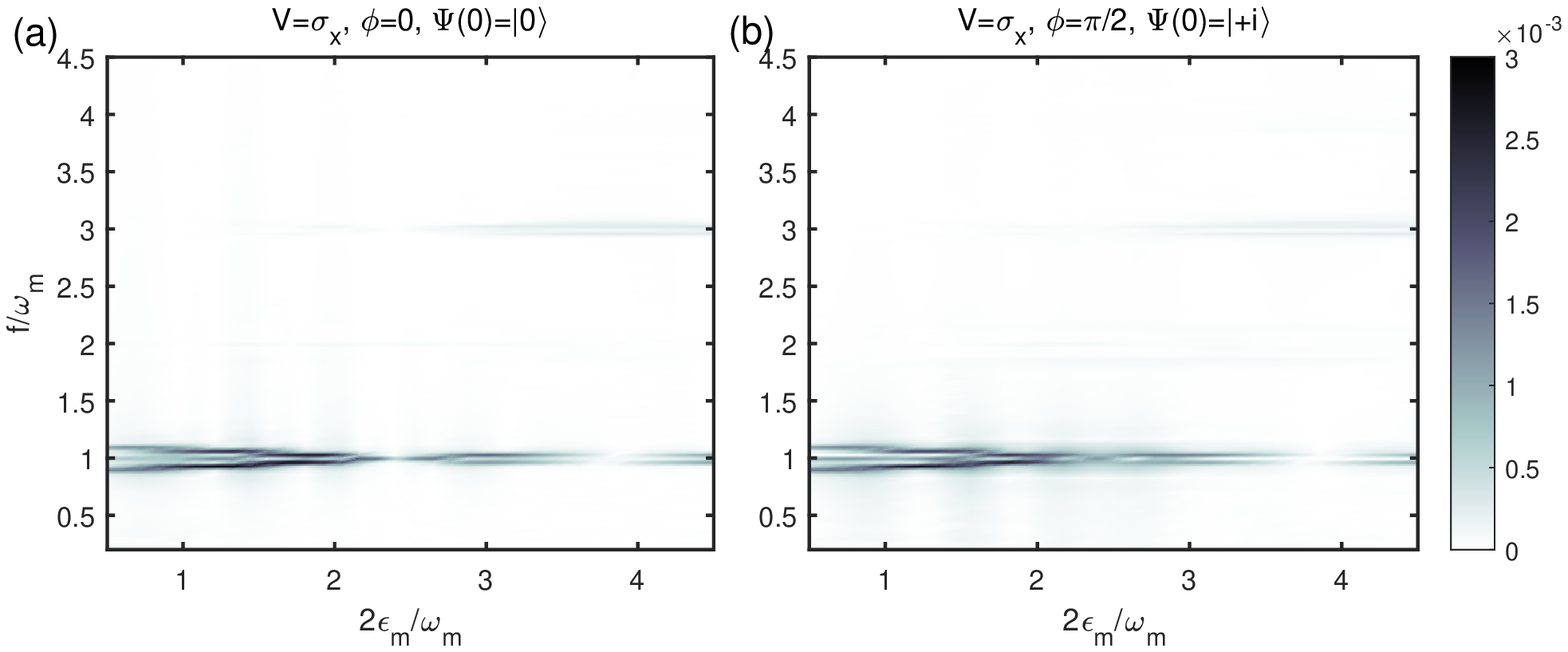}
\caption{\label{CDT_data_Supp} Experimental observation of siT, aDS, spDS, and spDB under a strong coupling regime. (a) Qubit initial state is prepared to $\ket{0}$ and the time-dependent population on $\ket{+}$ is readout (${V}=\sx$) from $t=0$ to $t=2\mu $s. Driving parameters are $\Omega=(2\pi)1.5\text{MHz}$, $\omega_m=(2\pi)15\text{MHz}$, $\phi=0$.  The intensity plot is the Fourier spectrum of measured Rabi oscillations under different $\epsilon_m$. (b) Same experiment as (a) except for the initial state $\ket{+i}$ (eigenstate of $\sy$) and the modulation phase $\phi=\pi/2$. }
\end{figure}
\begin{figure}[h]
\includegraphics[width=\textwidth]{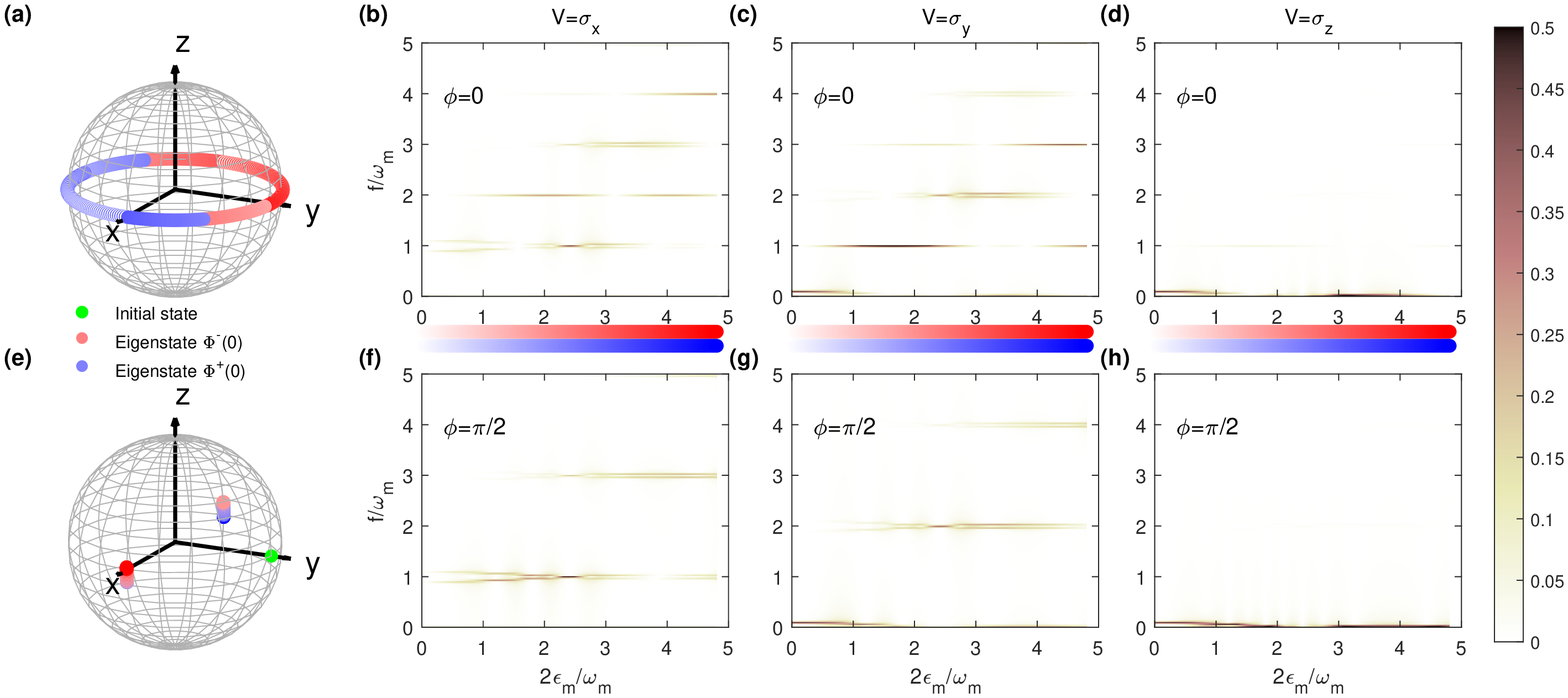}
\caption{\label{CDT_Simulation} Simulation of siT, aDS and spDS under a strong coupling regime. Qubit initial state is prepared to $\ket{0}$ and the Rabi oscillations are simulated from $t=0$ to $t=2\mu $s. Parameters are $\Omega=(2\pi)1.5\text{MHz}$, $\omega_m=(2\pi)15\text{MHz}$, $\phi=0,\pi/2$. (a,e) Visualization of Floquet eigenstates $\ket{\Phi^\pm(0)}$ on the Bloch sphere. The color change from white to red or blue as the increase of $2\epsilon_m/\omega_m$ from $0$ to $4.8096$. (b,f) Fourier spectrum of Rabi measurement on $\ket{+}$ (${V}=\sx$) under different modulation strength $2\epsilon_m/\omega_m$. (c,g) Fourier spectrum of Rabi measurement on $\ket{+i}$ (${V}=\sy$) under different modulation strength $2\epsilon_m/\omega_m$. (d,h) Fourier spectrum of Rabi measurement on $\ket{0}$ (${V}=\sz$) under different modulation strength $2\epsilon_m/\omega_m$.}
\end{figure}

Symmetry-induced transparency (siT) or coherent destruction of tunneling (CDT) are generated by destructive interference of two $V_{\mu,\nu}^{(n)}\neq 0$. In the context of light scattering,  siT requires two terms in the susceptibility in Eq.~\eqref{susceptibility} to cancel each other~\cite{engelhardt_dynamical_2021} when 
\begin{equation}
    V_{\mu^\prime,\mu}^{(-n-m)}V_{\mu,\mu^\prime}^{(m)}=V_{\mu,\mu^\prime}^{(-n-m)}V_{\mu^\prime,\mu}^{(m)}.
\end{equation}
In the context of coherent state evolution, the siT requires amplitudes of two sidebands to destructively interfere, such that
\begin{equation}
    c^+c^{-*}V_{-,+}^{(n)}+c^{+*}c^-V_{+,-}^{(n)}=0
\end{equation}
when they become degenerate at $\lambda^\pm=0$.

When $\Omega\ll\epsilon_m,\omega_m$, an additional particle-hole symmetry $\hat{P_2}=\hat{P_2}^\dagger=I$ is introduced. Combining with the always-on particle-hole symmetry $\hat{P_1}=\hat{P_1}^\dagger=\sz$, we denote them as $\hat{P}_1=\sz,\hat{P}_2=I$. Different from $\hat{P}_1$ that maps the Floquet eigenstate to its counterpart with opposite eigenenergy with $\ket{\Phi^{\mu^\prime}(t)}=\pi_\mu^{(P_1)}\ket{\Phi^{\mu *}(t+T/2)}$ where $\mu^\prime=\mp$ for $\mu=\pm$, $\hat{P}_2$ maps the Floquet eigenstate to itself with $\ket{\Phi^\mu(t)}=\pi_\mu^{(P_2)}\ket{\Phi^{\mu *}(t+T/2)}$ as shown in Fig.~\ref{Eigenstate_Symmetry_twoPH} (the more detailed derivations and explanations can be found in supplemental materials of Ref.~\cite{engelhardt_dynamical_2021}). These results can be summarized as 
\begin{align}
    \pi_\mu^{(P_1)}\hat{P}_1\ket{\Phi^{\mu*}(t+T/2)}&=\ket{\Phi^{\mu^\prime}(t)}\nonumber\\
    \pi_\mu^{(P_2)}\hat{P}_2\ket{\Phi^{\mu*}(t+T/2)}&=\ket{\Phi^\mu(t)}
\end{align}
where $\pi_\mu^{(P)}\pi_{\mu^\prime}^{(P)*}=1$, $\mu^\prime=\mp$ for $\mu=\pm$, as derived above in Sec.~\ref{Sec:Symmetries}. Then the dynamical dipole matrix element is 
\begin{align}
\label{Eq:derive_P2}
        V_{\mu,\nu}^{(n)}=&\frac{1}{T}\int_0^T \langle\Phi^\mu(t)|{V}|\Phi^\nu(t)\rangle e^{-in\omega_mt}dt=\frac{1}{T}\int_0^{T} \langle\Phi^{\mu *}(t+T/2)|\pi_{\mu}^{(P_2) *}\hat{P}_2^\dagger{V}\hat{P}_2\pi_{\nu}^{(P_2)}|\Phi^{\nu *}(t+T/2)\rangle  e^{-in\omega_mt}dt 
        \nonumber\\=&\alpha_V^{(P_2)}e^{i\pi n}V_{\mu,\nu}^{(-n) *}=\alpha_V^{(P_2)}e^{i\pi n}V_{\nu,\mu}^{(n)}
\end{align}
where $\alpha_V^{(P_2)}=1$ since the symmetry operator is identity.

In summary, two particle-hole symmetries result in the following relations
\begin{align}
    V_{\mu,\nu}^{(n)}&=\alpha_V^{(P_1)}e^{i\pi n}V_{\nu^\prime,\mu^\prime}^{(n)},\\
    V_{\mu,\nu}^{(n)}&=\alpha_V^{(P_2)}e^{i\pi n}V_{\nu,\mu}^{(n)}=e^{i\pi n}V_{\nu,\mu}^{(n)}.
\end{align}
As discussed in Sec.~\ref{Sec:Symmetries}, the selection rules for the sidebands induced by the first particle-hole symmetry $\hat{P}_1=\sz$ are already included in the parity symmetry prediction, and destructive (or constructive) interference happens for the centerbands when initial state is an equal superposition of Floquet eigenstates. Considering the second particle-hole symmetry $\hat{P}_2=I$, destructive (or constructive) interference happens in the sidebands when they have the same frequency (degenerate condition $\lambda^\pm=0$). (1) When $c^+c^{-*}=c^{+*}c^-$ (e.g., initial state $\ket{0}$, modulation phase $\phi=0$ or $\phi=\pi/2$), destructive interference happens for the odd bands with $e^{i\pi n}=-1$, which is observed in simulation shown in Figs.~\ref{CDT_Simulation}(b,f) where a clear siT happens at $2\epsilon_m/\omega_m=2.4048$. (2) When $c^+c^{-*}=-c^{+*}c^-$ (e.g., initial state $\ket{+i}$, modulation phase $\phi=\pi/2$), constructive interference happens for the odd bands with $e^{i\pi n}=-1$, which is observed in simulation shown in Figs.~\ref{CDT_Simulation_InitialY}(b,f).

We also perform experiments to observe both the destructive and constructive interferences with different qubit initial states (Fig.~\ref{CDT_data_Supp}.) When the qubit initial state is $\ket{0}$, such that $c^+c^{-*}=c^{+*}c^-$, and $V_{+,-}^{(n)}=-V_{+,-}^{(n)}$ at $2\epsilon_m/\omega_m=2.4048$, a destructive interference between two sidebands happens and siT or CDT are observed in Fig.~\ref{CDT_data_Supp}(a). However, when the qubit initial state is prepared to $\ket{+i}$ such that $c^+c^{-*}=-c^{+*}c^-$, a constructive interference happens at the degeneracy point as observed in Fig.~\ref{CDT_data_Supp}(b). 

In both derivations of Eq.~\eqref{Eq:derive_P1} and Eq.~\eqref{Eq:derive_P2}, the observation operator needs to satisfy $\hat{P}^\dagger{V}\hat{P}=\alpha_V^{(P)}{V}^*$. Thus, the previous derivations do not predict the situation of ${V}=\sy$. However, siT is also observed for ${V}=\sy$ as shown in simulation in Figs.~\ref{CDT_Simulation}(c,g). Then the question is, can the siT observed by $\sy$ also explained by the particle-hole symmetry? Following we will show that the answer is yes! 

Here we combine both two particle-hole symmetries in the derivation and calculate the dynamical dipole matrix element
\begin{align}
        V_{\mu,\nu}^{(n)}=&\frac{1}{T}\int_0^T \langle\Phi^\mu(t)|{V}|\Phi^\nu(t)\rangle e^{-in\omega_mt}dt=\frac{1}{T}\int_0^{T} \langle\Phi^{\mu *}(t+T/2)|\pi_{\mu}^{(P_2) *}\hat{P}_2^\dagger{V}\hat{P}_2\pi_{\nu}^{(P_2)}|\Phi^{\nu *}(t+T/2)\rangle  e^{-in\omega_mt}dt 
        \nonumber\\=&\frac{1}{T}\int_0^{T} \langle\Phi^{\mu^\prime}(t+T)|\hat{P}_1^{\dagger *}\pi_{\nu^\prime}^{(P_1)}\pi_{\mu}^{(P_2) *}\hat{P}_2^\dagger{V}\hat{P}_2\pi_{\nu}^{(P_2)}\pi_{\mu^\prime}^{(P_1)*}\hat{P}_1^*|\Phi^{\nu^\prime}(t+T)\rangle  e^{-in\omega_mt}dt \nonumber
        \\=&\alpha_V^{(P_1^*)}V_{\mu^\prime,\nu^\prime}^{(n)}
\end{align}
where $\pi_\mu^{(P)}\pi_{\nu}^{(P)*}=1,\hat{P}_2=I$ are used and $\alpha_V^{(P_1^*)}$ is given by $\alpha_V{V}=\hat{P}_1^{\dagger *}{V}\hat{P}_1^*$. Since $\hat{P}_1=\hat{P}_1^*=\sz$, then $\alpha_V^{(P_1^*)}=-1$ and $V_{+,-}^{(n)}=-V_{-,+}^{(n)}$ for both ${V}=\sx,\sy$ and destructive interference happens for both situations. Note that we do not require ${V}={V}^*$ in the derivation here.

\begin{figure}[h]
\includegraphics[width=\textwidth]{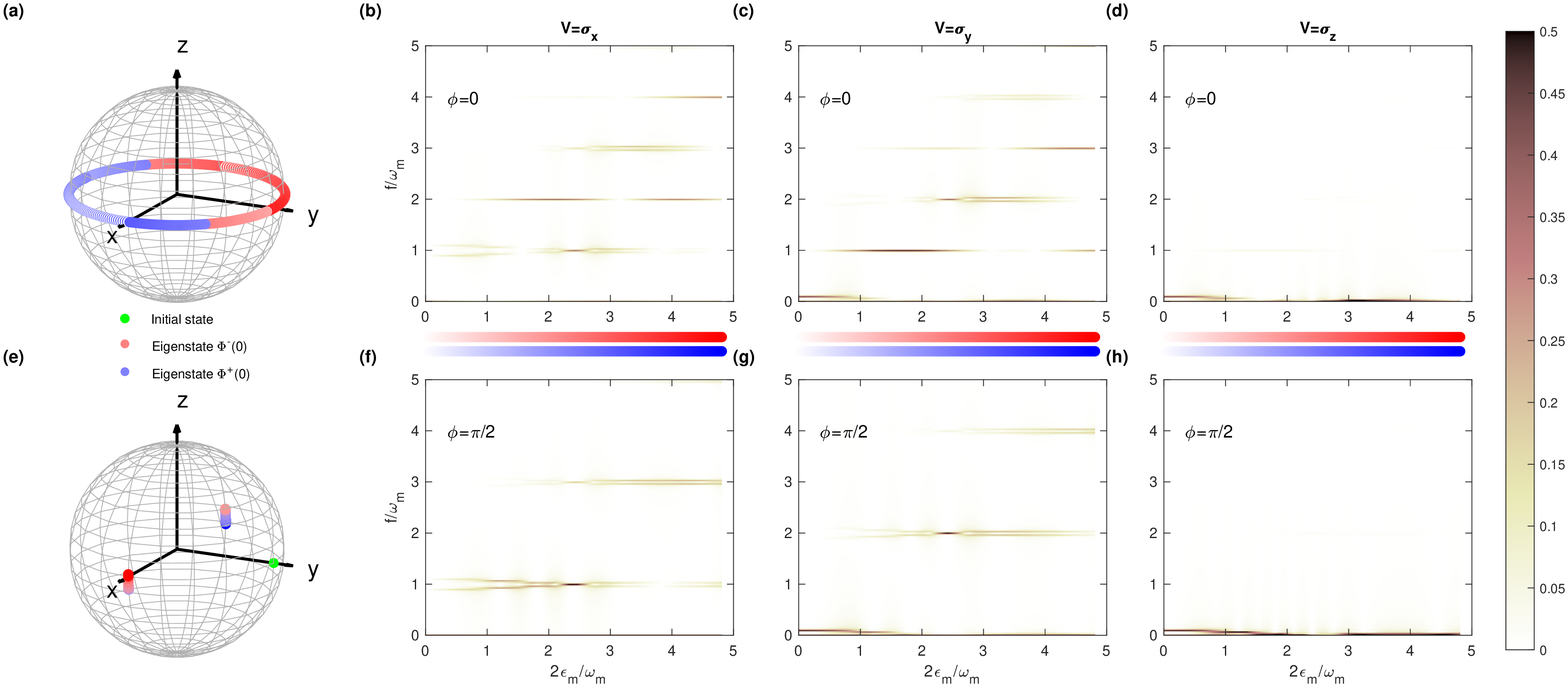}
\caption{\label{CDT_Simulation_InitialY} Simulation of siT, aDS, spDS and spDB under a strong coupling regime. Qubit initial state is prepared to $\ket{+i}$ (the eigenstate of $\sy$) and the Rabi oscillations are simulated from $t=0$ to $t=2\mu $s. Parameters are $\Omega=(2\pi)1.5\text{MHz}$, $\omega_m=(2\pi)15\text{MHz}$, $\phi=0,\pi/2$. (a,e) Visualization of Floquet eigenstates $\ket{\Phi^\pm(0)}$ on the Bloch sphere. The color change from white to red or blue as the increase of $2\epsilon_m/\omega_m$ from $0$ to $4.8096$. (b,f) Fourier spectrum of Rabi measurement on $\ket{+}$ (${V}=\sx$) under different modulation strength $2\epsilon_m/\omega_m$. (c,g) Fourier spectrum of Rabi measurement on $\ket{+i}$ (${V}=\sy$) under different modulation strength $2\epsilon_m/\omega_m$. (d,h) Fourier spectrum of Rabi measurement on $\ket{0}$ (${V}=\sz$) under different modulation strength $2\epsilon_m/\omega_m$.}
\end{figure}

Under the siT condition, the transition between $\ket{0}$ and $\ket{1}$ is suppressed, which is also called coherent destruction of tunneling. To further visualize such an interesting phenomena, we plot the state evolution on the Bloch sphere under four different conditions $\omega_m=2\Omega,6\Omega,10\Omega,20\Omega$ in Figs.~\ref{CDT_evolution}(a,b,c,d) with their corresponding Rabi oscillations in Figs.~\ref{CDT_evolution}(e,f,g,h). From the comparison, we can see that as the decrease of $\Omega/\omega_m$, the second particle-hole symmetry is gradually prominent, and the transition between $\ket{0}$ and $\ket{1}$ is more suppressed as predicted by the symmetry analysis.

\begin{figure}[h]
\includegraphics[width=\textwidth]{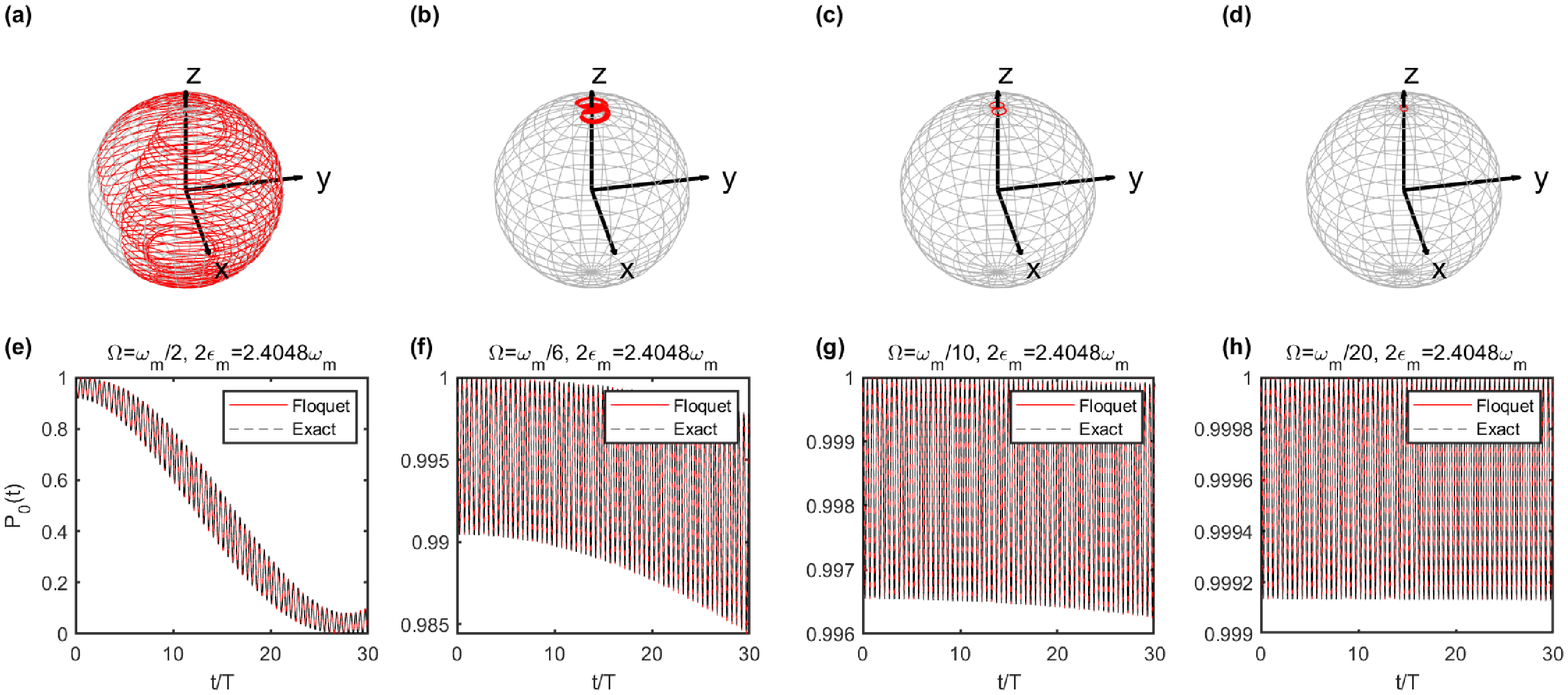}
\caption{\label{CDT_evolution} Simulation of siT (CDT) under different $\Omega$. Parameters are $\omega_m=(2\pi)15\text{MHz}$, $\phi=0, 2\epsilon_m/\omega_m=2.4048$ and the qubit evolutions of an initial state $\ket{0}$ are plotted on Bloch sphere in (a-d) and population on $\ket{0}$ is plotted in (e-h) (a,e) $\Omega=\omega_m/2$. (b,f) $\Omega=\omega_m/6$. (c,g) $\Omega=\omega_m/10$. (d,h) $\Omega=\omega_m/20$. }
\end{figure}

\clearpage
\section{S5. Rotation symmetry in a 3-level system}
\label{Sec:3-fold rotation symmetries}
\subsection{A. Hamiltonian engineering by amplitude or phase modulation}
In this section we introduce a novel method that exploits simultaneous modulated driving of various transitions  as a playground to study rotation symmetry. In particular, focusing on the NV center in diamond, here we show that even with two modulated microwave drivings, all  three energy levels can be coupled in the designed rotating frame.  

\textit{Amplitude modulation} - We start with the ground state Hamiltonian of NV center in the lab frame while simultaneously driving the transition between $\ket{m_s=0}$ and $\ket{m_s=\pm1}$.
\begin{equation}
H = \gamma BS_z + D S_z^2 + \sqrt{2}  \left[\Omega_1\cos(\omega_1t)-2\epsilon_{m1}(t)\sin(\omega_1t) + \Omega_2\cos(\omega_2t)+2\epsilon_{m2}(t) \sin(\omega_2t) \right]S_x
\end{equation}

We consider identical detuning for both transitions $\Delta = D\pm \gamma B -\omega_{1(2)}$. In the rotating frame defined by $U_1=diag(e^{-i\omega_1t},0,e^{-i\omega_2 t})$, under the rotating wave approximation we reach
\begin{equation}
\label{eq:H_first_frame}
\tilde{H} =
\begin{pmatrix}
   0 & \frac{\Omega_1}{2}- i \epsilon_{m1}(t)& 0\\
     \frac{\Omega_1}{2} +i \epsilon_{m1}(t) & \Delta &  \frac{\Omega_2}{2}- i \epsilon_{m2}(t) \\
     0 & \frac{\Omega_2}{2}- i \epsilon_{m2}(t) & 0 \\
    \end{pmatrix}
\end{equation}
where a constant term $diag(-\Delta,-\Delta,-\Delta)$ is neglected.
While the relation between detuning $\Delta$ and driving strength $\Omega_{1(2)}$ can be adjusted in a considerable range, we consider the case when $\Omega_1/2=\Omega_2/2=\Delta$. We then diagonalize the static part of Eq.~\ref{eq:H_first_frame}:
\begin{equation}
\tilde{H_T} =
\begin{pmatrix}
   2\Delta &  -i \frac{\epsilon_{m1}(t)+\epsilon_{m2}(t)}{\sqrt{3}}& i \frac{\epsilon_{m1}(t)-\epsilon_{m2}(t)}{\sqrt{2}}\\
     i \frac{\epsilon_{m1}(t)+\epsilon_{m2}(t)}{\sqrt{3}} & 0 & -i \frac{\epsilon_{m1}(t)+\epsilon_{m2}(t)}{\sqrt{6}} \\
     -i \frac{\epsilon_{m1}(t)-\epsilon_{m2}(t)}{\sqrt{2}} & i \frac{\epsilon_{m1}(t)+\epsilon_{m2}(t)}{\sqrt{3}} & -\Delta \\
    \end{pmatrix}.
\end{equation}
To reach the target Hamiltonian which has the same driving strengths but different phases between all the three energy levels, we set the time modulation to be
\begin{equation}
\epsilon_{m1}(t)=2\sqrt{3}J\sin(2\Delta t)\cos(\omega_m t+\frac{2\pi}{3})-2\sqrt{2}J\sin(3\Delta t)\cos(\omega_m t), \quad
\epsilon_{m2}(t)=2\sqrt{6}J\sin(\Delta t)\cos(\omega_mt+\frac{4\pi}{3}).
\end{equation}
Under the rotating frame defined by $U_2=diag(e^{-i2\Delta t},0,e^{i\Delta t})$, we finally get
\begin{equation}
\label{HI_Spin1}
\mathcal H_I =
\begin{pmatrix}
   0 & J\cos(\omega_m t+\frac{2\pi}{3}) &  J\cos(\omega_m t)\\
      J\cos(\omega_m t+\frac{2\pi}{3}) & 0 &  J\cos(\omega_m t+\frac{4\pi}{3}) \\
      J\cos(\omega_m t) &  J\cos(\omega_m t+\frac{4\pi}{3}) & 0 \\
    \end{pmatrix}.
\end{equation}
Note that here we use the rotating wave approximation which holds when $\Delta \gg \omega_m, J$ (see Fig.~\ref{Spin1_simulation_Supp}(c) for a symmetry breakdown caused by the dropped fast oscillating terms in this approximation). The time period of the above Hamiltonian is $T=2\pi/\omega_m$.

\textit{Phase modulation} - An alternative way to engineer a similar Hamiltonian is the phase modulation, where two phase modulated microwaves are applied to the transition between $\ket{m_s=0}$ and $\ket{m_s=\pm1}$.
\begin{equation}
H = \gamma BS_z + D S_z^2 + \sqrt{2}  \left[\Omega_1\cos(\omega_1t +f_1(t)) +\Omega_2\cos(\omega_2t +f_2(t)) \right]S_x
\end{equation}

We consider identical detuning for both transitions $\Delta = D\pm \gamma B -\omega_{1(2)}$. In the rotating frame defined by $U_1=diag(e^{-i(\omega_1t+f_1(t))},0,e^{-i(\omega_2t+f_2(t))})$, under the rotating wave approximation we reach
\begin{equation}
\label{eq:H_first_frame_phasemod}
\tilde{H} =
\begin{pmatrix}
   \epsilon_{m1}(t) & \frac{\Omega_1}{2}& 0\\
     \frac{\Omega_1}{2}  & \Delta &  \frac{\Omega_2}{2} \\
     0 & \frac{\Omega_2}{2} & \epsilon_{m2}(t) \\
    \end{pmatrix}
\end{equation}
where a constant term $diag(-\Delta,-\Delta,-\Delta)$ is neglected, and $\epsilon_{1(2)}=-\frac{d}{dt}f_{1(2)}(t)$.
While the relation between detuning $\Delta$ and driving strength $\Omega_{1(2)}$ can be adjusted in a considerable range, we consider the case when $\Omega_1/2=\Omega_2/2=\Delta$. We then diagonalize the static part of Eq.~\ref{eq:H_first_frame_phasemod}:
\begin{equation}
\tilde{H_T} =
\begin{pmatrix}
   2\Delta +\frac{\epsilon_{m1}(t)+\epsilon_{m2}(t)}{6} &  \frac{-\epsilon_{m1}(t)+\epsilon_{m2}(t)}{2\sqrt{3}}& \frac{\epsilon_{m1}(t)+\epsilon_{m2}(t)}{3\sqrt{2}}\\
     \frac{-\epsilon_{m1}(t)+\epsilon_{m2}(t)}{2\sqrt{3}} & \frac{\epsilon_{m1}(t)+\epsilon_{m2}(t)}{2} & \frac{-\epsilon_{m1}(t)+\epsilon_{m2}(t)}{\sqrt{6}} \\
     \frac{\epsilon_{m1}(t)+\epsilon_{m2}(t)}{3\sqrt{2}} & \frac{-\epsilon_{m1}(t)+\epsilon_{m2}(t)}{\sqrt{6}} & -\Delta+\frac{\epsilon_{m1}(t)+\epsilon_{m2}(t)}{3} \\
    \end{pmatrix}.
\end{equation}
To reach the target Hamiltonian which has the same driving strengths but different phases between all the three energy levels, we set the modulation to be
\begin{equation}
\epsilon_{m1}(t)=-4\sqrt{3}J\sin(2\Delta t)\cos(\omega_m t+\frac{2\pi}{3})+6\sqrt{2}J\sin(3\Delta t)\cos(\omega_m t), \quad
\epsilon_{m2}(t)=2\sqrt{6}J\sin(\Delta t)\cos(\omega_mt+\frac{4\pi}{3}).
\end{equation}
Under the rotating frame defined by $U_2=diag(e^{-i2\Delta t},0,e^{i\Delta t})$, we finally get the same Hamiltonian as in Eq.~\eqref{HI_Spin1}. Note that here we also use the similar rotating wave approximation.

\subsection{B. Symmetry protected selection rules}

\begin{figure}[h]
\includegraphics[width=0.8\textwidth]{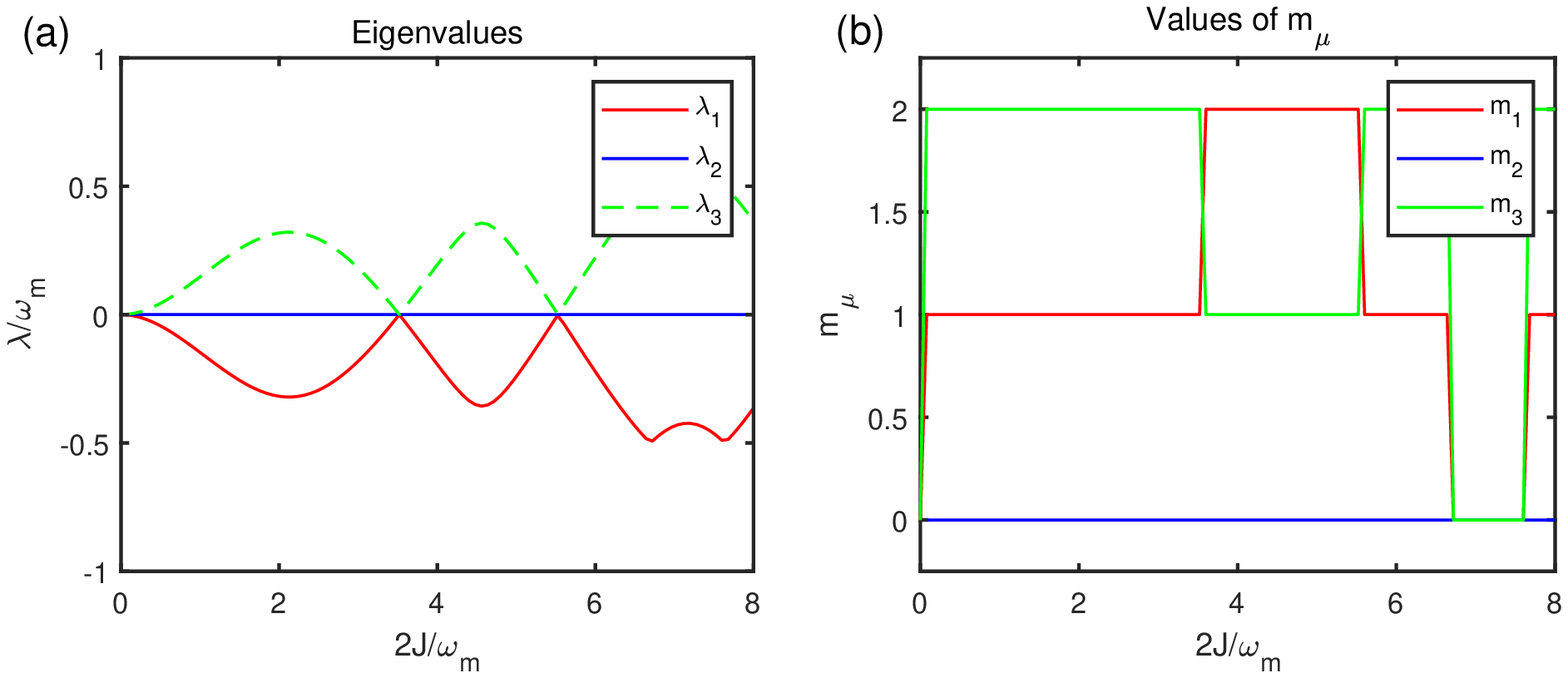}
\caption{\label{Eigenstate_Symmetry_Spin1_Supp} Floquet eigenenergies and $\pi_\mu$. (a) Floquet eigenenergies $\lambda^{1,2,3}$. Note that $\lambda$ is defined in the range $(-\omega_m/2,\omega_m/2]$ and $\lambda^1\leq\lambda^2\leq\lambda_3$. (b) The values of $m_{1,2,3}$ corresponding to the rotation symmetry. Floquet simulations in this figure are implemented under the condition $\omega_m=(2\pi)0.3\text{MHz}$. }
\end{figure}

\begin{figure}[h]
\includegraphics[width=\textwidth]{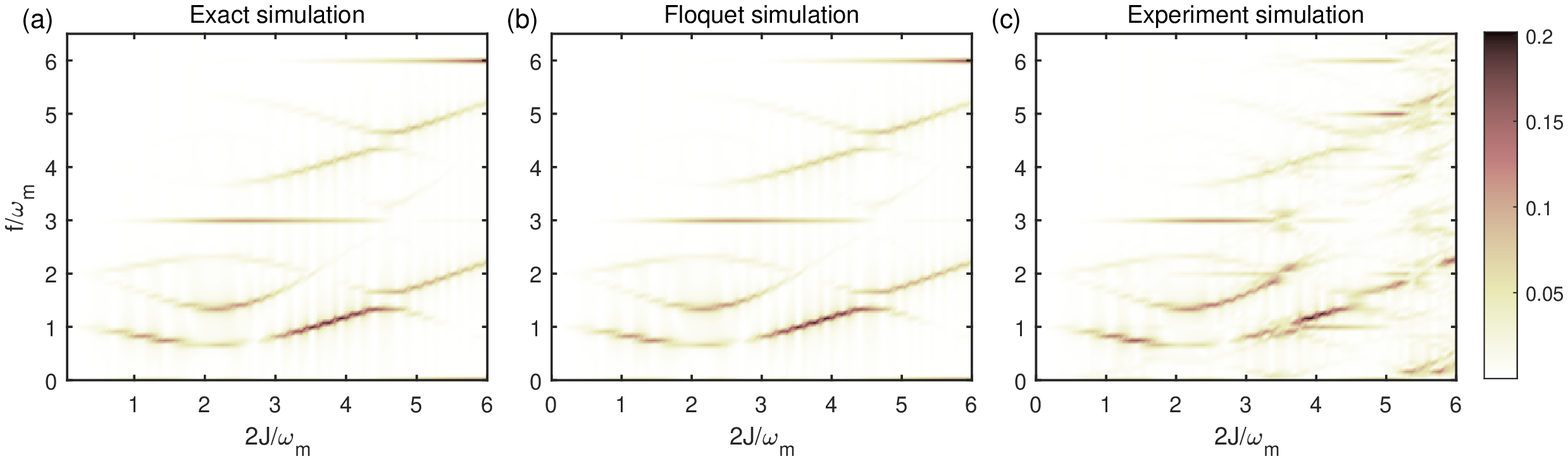}
\caption{\label{Spin1_simulation_Supp} Simulation of spDSs and spDBs in a 3-level system. The initial state is prepared to $(1/\sqrt{3})(\ket{e_1}+\ket{e_2}+\ket{e_3})$ where $\ket{e_1}=(1/\sqrt{3})[1,1,1]^T$, $\ket{e_2}=(1/\sqrt{6})[-2,1,1]^T$, $\ket{e_3}=(1/\sqrt{2})[0,1,-1]^T$ are three eigenstates of ${V}$, such that evolution mode involves all bands. Driving parameters $\omega_m=(2\pi)0.3\text{MHz}$, and $2J/\omega_m$ is swept from $0$ to $6$. The intensity plot is the Fourier spectrum of simulated Rabi oscillations under different $2J/\omega_m$. (a) Exact simulation. The time-dependent weighted Rabi population $P(t)$ is simulated from $t=0$ to $t=40\mu $s with 5001 sampling points. Hamiltonian in Eq.~\eqref{HI_Spin1} is used to calculate the exact evolution. (b) Floquet simulation. The Floquet matrix is truncated to a 150 blocks by 150 blocks, the evolution time is swept from $t=0$ to $t=40\mu$s with 5001 sampling points. (c) Experiment simulation. Hamiltonian in the first rotating frame in Eq.~\eqref{eq:H_first_frame} is used where $\Omega_1=\Omega_2=2\Delta=(2\pi)30\text{ MHz}$ is used. The time-dependent weighted Rabi population $P(t)$ is simulated from $t=0$ to $t=40\mu $s with 160001 sampling points to achieve high precision. }
\end{figure}

The symmetry associated with the Hamiltonian $\ham_I$ in Eq.~\eqref{HI_Spin1} is a 3-fold rotation symmetry
\begin{equation}
    \hat{R} = \begin{pmatrix}
    0 & 0 & 1\\
    1 & 0 & 0\\
    0 & 1 & 0\\
    \end{pmatrix},
\end{equation}
which simply gives $\hat{R}\ham_I(t+T/3)\hat{R^\dagger}=\ham_I(t)$. Such a symmetry gives rise to a relation $\ket{\Phi^\mu(t)}=\pi_\mu^{(R)}\hat{R}\ket{\Phi^\mu(t+T/3)}=e^{i2\pi m_\mu /3}\hat{R}\ket{\Phi^\mu(t+T/3)}$ with $\mu\in\{1,2,3\}$ and $m_\mu\in\{0,1,2\}$~\cite{engelhardt_dynamical_2021}. 
The values of $\pi_\mu^{(R)}=e^{i2\pi m_\mu /3}$ have dependence on the driving parameter $J$ (also on how we define the bands). In Figs.~\ref{Eigenstate_Symmetry_Spin1_Supp}(b), we simulate and plot the three values of $m_{1,2,3}$ as a dependence of $2J/\omega_m$ where $m_{1,3}$ switch with each other when their eigenvalues cross the degeneracy point. 

Now we start to derive the selection rule by evaluating the value of dynamical dipole matrix element. The dynamical dipole matrix element is
\begin{align}
    V_{\mu,\nu}^{(n)}&=\frac{1}{T}\int_0^T \langle\Phi^\mu(t)|{V}|\Phi^\nu(t)\rangle e^{-in\omega_mt}dt\nonumber\\&=\frac{1}{T}\int_0^{T/3} \langle\Phi^\mu(t)|{V}|\Phi^\nu(t)\rangle e^{-in\omega_mt}dt\times\left[1+\pi_\mu^{(R)}\pi_\nu^{(R)*}e^{-i\frac{2\pi}{3} n}\alpha_V+\left(\pi_\mu^{(R)}\pi_\nu^{(R)*}e^{-i\frac{2\pi}{3} n}\alpha_V\right)^2\right]
\end{align}
where the value of $\alpha_V$ is given by relation $\hat{R}{V}\hat{R}^\dagger=\alpha_V{V}$. Since $\alpha_V$, $\pi_{\mu}^{(R)}$ and $\pi_{\nu}^{(R)}$ can only take values from $\{e^{i\frac{2\pi}{3}},e^{i\frac{4\pi}{3}},1\}$ due to their 3-fold rotational symmetry, the band vanishing condition is that $e^{i\frac{2\pi}{3} (m_\mu-m_\nu-n)}\alpha_V\neq 1$, which is validated in the simulation shown in the main text combining with supplementary Figs.~\ref{Eigenstate_Symmetry_Spin1_Supp}.

In this work, we evaluate the observation operator 
\begin{equation}
    {V} = \begin{pmatrix}
    0 & 1 & 1\\
    1 & 0 & 1\\
    1 & 1 & 0\\
    \end{pmatrix},
\end{equation}
which gives $\alpha_V=1$. This observation operator can be rewritten in its eigen basis as $V=2\ket{e_1}\bra{e_1}-\ket{e_2}\bra{e_2}-\ket{e_3}\bra{e_3}$ where $\ket{e_1}=(1/\sqrt{3})[1,1,1]^T$, $\ket{e_2}=(1/\sqrt{6})[-2,1,1]^T$, $\ket{e_3}=(1/\sqrt{2})[0,1,-1]^T$. Thus, the weighted Rabi signal is defined as $P(t)=(1/4)\left[2P_{\ket{e_1}}(t)-P_{\ket{e_2}}(t)-P_{\ket{e_3}}(t)\right]$. The Fourier spectrum of this signal presenting the spDSs and spDBs is shown in both the main text and in Fig.~\ref{Spin1_simulation_Supp}.

To further mimic a practical experiment, we perform the exact simulation in the first rotating frame [Eq.~\eqref{eq:H_first_frame}] where fast oscillating terms at harmonics of frequency $\Delta$ are not neglected as in Eq.~\eqref{HI_Spin1}. We take the parameters $\omega_m=(2\pi)0.3\text{ MHz}$, $\Delta=\Omega_1/2=\Omega_2/2=(2\pi)15\text{ MHz}$, and show the result in Fig.~\ref{Spin1_simulation_Supp}(c). 
In comparison to both exact simulation and Floquet simulation in Figs.~\ref{Spin1_simulation_Supp}(a,b), some protected spDSs or spDBs start to appear at larger $J$ where the rotating wave approximation starts to break down.

\clearpage
\section{S6. Experimental imperfections}\label{sec:experiment}
In the comparison between experiment in Fig.~\ref{Resonance_data_Supp} and exact simulations in Fig.~\ref{Resonance_Simulation}, we find that some unexpected centerbands are observed experimentally, which are forbidden by the selection rules. Since we are using a qubit ensemble with $10^{10}$ qubits being addressed simultaneously, the dominant noise comes from the inhomogeneities as discussed in detail in Ref.~\cite{wangCoherenceProtectionDecay2020}. One possible reason for these unexpected bands is that the inhomogeneities introduce a detuning term $(\delta/2)\sz$ in the Hamiltonian $\ham_I$, which breaks both  parity and particle-hole symmetries. In Fig.~\ref{ResonanceInhomo_Simulation}, we simulate this effect assuming a normal distribution of the detuning term $\delta\sim\mathcal{N}(0,(2\pi)0.15\text{MHz})$, and the simulation shows the appearance of the unexpected centerbands and also sidebands. 
Since the centerband is highly robust against experimental noise and other imperfections as studied in Ref.~\cite{wangCoherenceProtectionDecay2020}, the unexpected centerbands are more prominent than the sidebands. 
 
Another possible reason for the breaking of the symmetry is the imperfect polarization of NV nuclear spin which includes three hyperfine levels corresponding to the three spin states of \Nit ($m_I=0,\pm1$) separated by the hyperfine coupling constant $A=(2\pi)2.16$MHz. 
Under the experimental condition when the magnetic field is $239\text{G}$, only $\sim 70\%$ of the NV spin is polarized to the hyperfine level $m_I=+1$ (which is the state we assume in this work), and there are still $30\%$ NV spins with detunings $A=(2\pi)2.16$MHz, $2A=(2\pi)4.32$MHz. These static detunings break the same dynamical symmetries and could also induce the unexpected bands.

A third  experimental imperfection is due to the nonlinearity of the microwave amplifier. In the supplemental materials of Ref.~\cite{wang_observation_2021}, we characterize the nonlinearity of the microwave amplifier in the same experimental setup and show that such nonlinearity may induce higher harmonics which affect the practically engineered Hamiltonian.

A fourth  experimental imperfection arises from the fact that the microwave also has longitudinal component which gives rise to an additional $\sigma_z$ term in the lab frame Hamiltonian that breaks the discussed symmetries.

Finally, the Hamiltonian $\ham_I$ was derived under the rotating wave approximation. For strong driving,   counter-rotating terms may also induce the symmetry breaking, which is validated in the comparison in Fig.~\ref{Spin1_simulation_Supp} where the simulation in the first rotating frame in Fig.~\ref{Spin1_simulation_Supp}(c) brings symmetry breaking comparing to the simulation in the second rotating frame in Fig.~\ref{Spin1_simulation_Supp}(a) and (b) due to the existence of the counter-rotating terms.
\begin{figure}[h]
\includegraphics[width=0.95\textwidth]{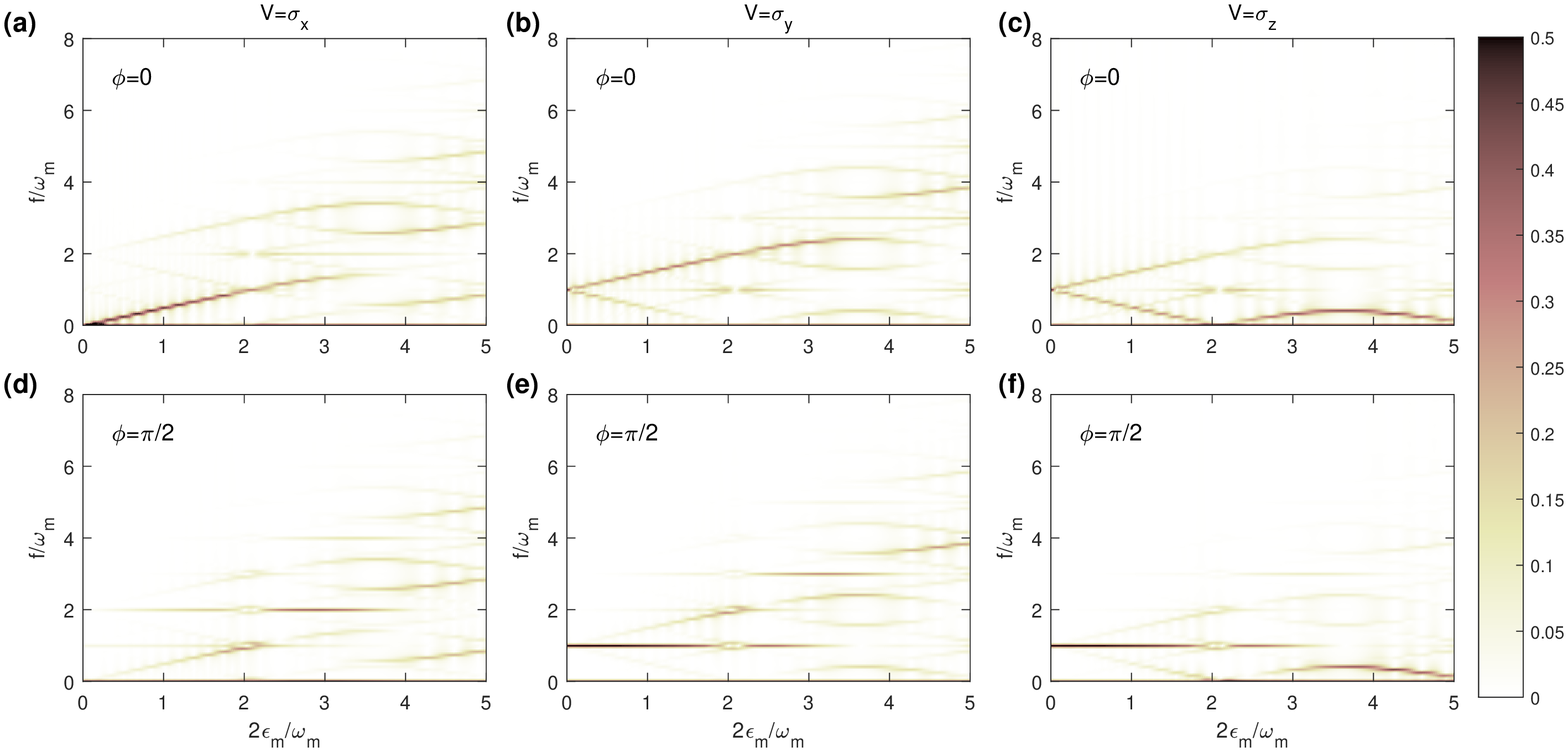}
\caption{\label{ResonanceInhomo_Simulation} Simulation of symmetry breaking due to inhomogeneities. Parameters are the same as in  Fig.~\ref{Resonance_Simulation} except for assuming a Normal distribution of the detuning $\delta\sim\mathcal{N}(0,\sigma=(2\pi)0.15\text{MHz})$ in Hamiltonian $\ham_I=(\delta/2)\sz+(\Omega/2)\sx+\epsilon_m\sin(\omega_mt+\phi)\sz$. To simplify the simulation, we sample 51 points from $\delta=-2\sigma$ to $\delta=2\sigma$ and calculate their average Rabi signals.}
\end{figure}

\clearpage

\end{widetext}

\end{document}